\documentclass[journal]{IEEEtran}
\usepackage{amsmath,amsfonts,amssymb}
\usepackage{multirow}
\usepackage{booktabs}
\usepackage{algorithmic}
\usepackage{algorithm}
\usepackage{array}
\usepackage[justification=centering,caption=false,font=normalsize,labelfont=sf,textfont=sf]{subfig}
\usepackage{textcomp}
\usepackage{stfloats}
\usepackage{url}
\usepackage{bbm}
\usepackage{comment}
\usepackage{verbatim}
\usepackage{graphicx}
\usepackage{cite}
\usepackage{xcolor}
\usepackage{tikz}
\usetikzlibrary{matrix}
\newtheorem{example}{Example}
\newtheorem{theorem}{Theorem}
\newtheorem{definition}{Definition}
\newtheorem{lemma}{Lemma}
\newtheorem{construction}{Construction}
\newtheorem{remark}{Remark}
\newtheorem{observation}{Observation}
\newtheorem{corollary}{Corollary}

\title{Asymptotically Optimal Codes for $(t,s)$-Burst Error}

\author{Yubo~Sun, Ziyang~Lu, Yiwei~Zhang, and Gennian~Ge%
\thanks{This work was supported in part by the National Key Research and Development Program of China under Grant 2020YFA0712100, in part by the National Natural Science Foundation of China under Grant 12231014, and in part by the Beijing Scholars Program. The research of Ziyang Lu and Yiwei Zhang was also supported in part by the National Key Research and Development Program of China under Grant 2022YFA1004900 and Grant 2021YFA1001000, in part by the Shandong Provincial Natural Science Foundation under Grant No. ZR2021YQ46, and in part by the Taishan Scholars Program.}
\thanks{Y. Sun and G. Ge are with the School of Mathematical Sciences, Capital Normal University, Beijing 100048, China (e-mail: 2200502135@cnu.edu.cn, gnge@zju.edu.cn). Z. Lu and Y. Zhang are with the Key Laboratory of Cryptologic Technology and Information Security, Ministry of Education, and also with the School of Cyber Science and Technology, Shandong University, Qingdao, Shandong 266237, China (e-mail: zylu@mail.sdu.edu.cn, ywzhang@sdu.edu.cn).}
}

\begin{document}
\maketitle

\begin{abstract}

Recently, codes for correcting a burst of errors have attracted significant attention.
One of the most important reasons is that bursts of errors occur in certain emerging techniques, such as DNA storage.
In this paper, we investigate a type of error, called a $(t,s)$-burst, which
deletes $t$ consecutive symbols and inserts $s$ arbitrary symbols at the same coordinate.
Note that a $(t,s)$-burst error can be seen as a generalization of a burst of insertions ($t=0$), a burst of deletions ($s=0$), and a burst of substitutions ($t=s$).
Our main contribution is to give explicit constructions of $q$-ary $(t,s)$-burst correcting codes with $\log n + O(1)$ bits of redundancy for any given constant non-negative integers $t$, $s$, and $q \geq 2$.
These codes have optimal redundancy up to an additive constant.
Furthermore, we apply our $(t,s)$-burst correcting codes to combat other various types of errors and improve the corresponding results. In particular, one of our byproducts is a permutation code capable of correcting a burst of $t$ stable deletions with $\log n + O(1)$ bits of redundancy, which is optimal up to an additive constant.
\end{abstract}


\section{Introduction}

DNA storage is a promising direction for future data storage. Unlike other storage systems, not only substitutions, but also insertions and deletions are common errors in DNA storage \cite{Church-12-science-DNA,Yazdi-15-TMBMC-DNA,Lee-19-NC-DNA,Heckel-19-background}.
Thus, investigating the constructions of codes that are capable of correcting insertions and deletions is of great significance. However, this is a notoriously challenging problem, partly because of the fact that the erroneous coordinates may still be unknown even if the correct sequence is known.

In 1966, Levenshtein \cite{Levenshtein-66-SPD-1D} first studied insertions and deletions errors and showed the equivalence between $t$-insertion correcting codes and $t$-deletion correcting codes, here and elsewhere the equivalence between two types of codes signifies that they have the same error correction capability, although the decoding procedures for addressing these two types of errors may differ.
Most importantly, he proved that the optimal redundancy of $t$-deletion correcting codes is lower bounded by $(t+o(1)) \log n$.
So far, constructions of $t$-deletion correcting codes achieving the lower bound exist only when $t=1$ \cite{Levenshtein-66-SPD-1D,Tenengolts-84-IT-q_D,Nguyen-22-arXiv-q_D}.
For $t=2$, Guruswami and H{\aa}stad \cite{Guruswami-21-IT-2D} presented an explicit construction with redundancy $(4+o(1))\log n$ for the binary case,
and for the non-binary case Song and Cai \cite{Song-22-arXiv-BD} and  Liu \emph{et al.} \cite{Liu-23-arXiv-2D} constructed codes with redundancy $(5+o(1)) \log n$ when the alphabet size satisfies certain conditions. 
For $t\geq3$, the existential best known results have redundancy $(4t-1+o(1)) \log n$ \cite{Song-22-IT-DS} and $(4t+o(1)) \log n$ \cite{Sima-20-ISIT-q-tD} for the binary and non-binary cases, respectively.

Both in DNA storage and some other communication systems, errors occur in bursts and tend to cluster together \cite{Schoeny-17-IT-BD,Chee-20-IT-PBSD,Chee-18-IT-RM,Sun-22-IT-BD,Sun-23-IT-BDR}.
The study of codes against bursts of deletions or insertions can be found in the following recent works \cite{Cheng-14-ISIT-BD,Gabrys-18-IT-D_T,Lenz-20-ISIT-BD,Levenshtein-70-BD,Schoeny-17-IT-BD,Wang-22-arXiv-BD,Saeki-18-ISIT-BD}.
While one can simply use the aforementioned $t$-deletion correcting codes to combat a burst of $t$ consecutive deletions (a $t$-burst-deletion), it is not the best choice since the erroneous coordinates are arbitrary in the former whereas concentrated in the latter. Taking advantage of this special feature, Schoeny \emph{et al.} \cite{Schoeny-17-IT-BD} derived that the optimal redundancy of binary $t$-burst-deletion correcting codes is lower bounded by $\log n+O(1)$ and showed the equivalence between $t$-burst-deletion correcting codes and $t$-burst-insertion correcting codes. Later Wang \emph{et al.} \cite{Wang-22-arXiv-BD} generalized their results to the non-binary alphabet.
By now, for $t\geq 2$ and for both the binary and non-binary cases, the existential best known $t$-burst-deletion correcting codes have redundancy $\log n + O( \log \log n)$ \cite{Schoeny-17-IT-BD,Saeki-18-ISIT-BD,Wang-22-arXiv-BD}, except for the only known asymptotically optimal binary $2$-burst-deletion correcting code with $\log n+O(1)$ bits of redundancy \cite{Levenshtein-70-BD}.

The terminology of burst-substitution error needs some clarifications.
Unlike burst-insertion and burst-deletion, in a $t$-burst-substitution error the erroneous coordinates are localized in an interval of length $t$ but not necessarily consecutive, i.e., within an interval of length $t$ some $t'\leq t$ coordinates are erroneous while the others could be {\it error free} \cite{Fire-59-BS, Chien-69-IT-BS,Zhou-16-IT-BS}.
For example, $001010$ can be obtained from $011110$ by a $3$-burst-substitution at the second coordinate (substituting $111$ with $010$), where the third coordinate is error free.
This type of error could be found in many applications, including high-density magnetic recording channels \cite{Kuznetsov-93-IT}, \cite{Levenshtein-93-IT}, flash memories \cite{Cassuto-10-IT}, and some DNA-based storage systems \cite{Jain-20-ISIT}.
Fire \cite{Fire-59-BS} presented binary burst-substitution correcting codes with $\log n+O(1)$ bits of redundancy.
Abdel-Ghaffar \emph{et al.} \cite{Abdel-Ghaffar-86-IT-BS, Abdel-Ghaffar-88-IT-BS} and
Etzion \cite{Etzion-01-IT-BS} studied perfect burst-substitution correcting codes under certain parameters. Recently Nguyen \textit{et al}. \cite{Nguyen-22-ISIT} and then later Wei and Schwartz \cite{Wei-23-IT-BS} considered several special types of bursts of substitutions.

In current DNA storage technology, one needs to design error-correcting codes combating a combination of insertions, deletions, and substitutions \cite{Heckel-19-background}, which is more difficult than combating a single type of error alone. The objective of this paper is to study codes against bursts of the three types of errors. Note that a $t$-burst-substitution error can be viewed as a $t$-burst-deletion error followed by a $t$-burst-insertion error at the same coordinate, and thus we mainly focus on bursts of insertions and deletions.

In this paper, we consider the $t$-deletion-$s$-insertion-burst error, abbreviated as a $(t,s)$-burst error, which deletes $t$ consecutive symbols and inserts $s$ arbitrary symbols at the same coordinate.
This type of error can be seen as a generalization of a burst of insertions when $t=0$, a burst of deletions when $s=0$, and a burst of substitutions when $t=s$.
Codes capable of correcting this type of error, the $(2,1)$-burst error to be specific, were first studied by Schoeny \emph{et al.} and their original motivation is to use $(2,1)$-burst correcting codes to deal with localized deletion errors \cite{Schoeny-17-IT-BD}. Later, Lu and Zhang \cite{Lu-22-arxiv-BD,Lu-22-ISIT-BD} formalized the $(t,s)$-burst error and showed the equivalence between $(t,s)$-burst correcting codes and $(s,t)$-burst correcting codes, and thus it suffices to only consider $t \geq s$. In \cite{Lu-22-arxiv-BD} it is also shown that the optimal redundancy of binary $(t,s)$-burst correcting codes is lower bounded by $\log n +O(1)$, where the constant depends only on $t$.
As for the constructions, $(t,s)$-burst correcting codes with redundancy $(1+o(1)) \log n$ are known only for the binary case and only when $t \geq 2s$ \cite{Schoeny-17-IT-BD,Lu-22-arxiv-BD,Lu-22-ISIT-BD}, see Table \ref{tab:r1} for details.
Moreover, the current best constructions have redundancy achieving the lower bound, up to an additive constant, only when $(t,s) \in \{(2,1),(3,1)\}$ \cite{Schoeny-17-IT-BD,Lu-22-arxiv-BD,Lu-22-ISIT-BD}.

Our main contribution is to construct $(t,s)$-burst correcting codes with $\log n +O(1)$ bits of redundancy for any alphabet size $q$ and any given constant non-negative integers $t$ and $s$. Furthermore, we use these codes to combat other types of errors and improve the redundancy under certain parameters.
Particularly, we construct a permutation code that is capable of correcting a burst of $t$ stable deletions with $\log n + O(1)$ bits of redundancy, and there is only a constant gap between the redundancy of our permutation code and the theoretic bound. We summarize our main results in Tables \ref{tab:r1} and \ref{tab:r2}.

\begin{table*}
\renewcommand\arraystretch{1.5}
    \caption{Results of $q$-ary $(t,s)$-burst correcting codes where $t \geq s$}
    \centering
    \begin{tabular}{cccc}
        \hline
        \hline
        ~ & Parameters $(q,t,s)$ & Code Redundancy & Work \\
        \hline

        \multirow{8}{*}{Previous results} & $(2,1,0)$ and $(2,1,1)$ & $\log n +O(1)$ & Levenshtein \cite{Levenshtein-66-SPD-1D} \\

        & $(2,2,0)$ & $\log n +O(1)$ & Levenshtein \cite{Levenshtein-70-BD} \\

        & $(2,2,1)$ & $\log n +O(1)$ & Schoeny \emph{et al.} \cite{Schoeny-17-IT-BD} \\

        & $(2,3,1)$ & $\log n +O(1)$ & Lu and Zhang \cite{Lu-22-arxiv-BD} \\

        & $(2,t,0)$ & $\log n +O(\log \log n)$ & Schoeny \emph{et al.} \cite{Schoeny-17-IT-BD} \\

        & $(2,t,s)$ for $t \geq 2s$ & $\log n +O(\log \log n)$ & Lu and Zhang \cite{Lu-22-arxiv-BD} \\

        & $(q,1,0)$ & $\log n +O(1)$ & Tenengolts \cite{Tenengolts-84-IT-q_D}\\

        & $(q,t,0)$ & $\log n +O(\log \log n)$ & Saeki and Nozaki \cite{Saeki-18-ISIT-BD}\\
        \hline

        Our results & $(q,t,s)$ & $\log n +O(1)$ & Theorems \ref{thm:(t,t)-burst} and \ref{thm:(t,s)-burst} \\
        \hline
        \hline
    \end{tabular}
    \label{tab:r1}
\end{table*}

\begin{table*}
\renewcommand\arraystretch{1.5}
    \caption{Applications of $(t,s)$-burst correcting codes}
    \centering
    \begin{tabular}{ccc}
        \hline
        \hline
        Applications & Previous Best Known Results & Our Results \\
        \hline

        $t$-inversion correcting codes &  $\log n + 4t \log q + 2$ \cite{Nguyen-22-ISIT} &  $\log n + 4(t-1) \log q + 2- \log(t-1)$ (Theorem \ref{thm:(t,t)-burst})  \\

        $^{\leq}t$-inversion correcting codes & $\log n + 7(t + 1) \log q + 4$ \cite{Nguyen-22-ISIT} & $\log n + 4(t-1) \log q + 2 - \log(t-1)$ (Theorem \ref{thm:(t,t)-burst})  \\

        Type-A-Absorption correcting code &  $\log n + 12 \log \log n + O(1)$ \cite{Ye-24-IT-Absorption} &  $\log n + O(1)$ (Theorem \ref{thm:(t,t)-burst})  \\

        Type-B-Absorption correcting code & N/A  &  $\log n + O(1)$ (Theorem \ref{thm:(t,s)-burst})  \\

        $^{\leq}t$-burst-deletion correcting codes & $\log n + \min \{\binom{t+1}{2},8\lceil \log q \rceil\} \log \log n + O(1)$ \cite{Lenz-20-ISIT-BD,Wang-22-arXiv-BD} & $\log n + t \log \log n + O(1)$ (Theorem \ref{thm:leq_t-burst}) \\

        $t$-localized-deletion correcting codes & $\log n+ 16t\log \log n +O(1)$ \cite{Bitar-21-ISIT-LD} & $\log n+ 2t\log \log n + O(1)$ (Theorem \ref{thm:localized_deletion}) \\

        $t$-BSD correcting permutation codes & $\log n + 2\log \log n+O(1)$ \cite{Sun-22-IT-BD} & $\log n + O(1)$ (Theorem \ref{thm:t-BSD_permutation_code}) \\

        $^{\leq}t$-BSD correcting permutation codes & $\log n + (3t-2)\log \log n+O(1)$ \cite{Sun-22-IT-BD} & $\log n + t \log \log n + O(1)$ (Theorem \ref{thm:leqt-BSD_permutation_code}) \\
        \hline
        \hline
    \end{tabular}
    \label{tab:r2}
\end{table*}

The rest of this paper is organized as follows.
Section \ref{Sec:Pre} introduces the relevant notations used throughout the paper.
Section \ref{Sec:Bound} calculates the sphere packing bound on the size of any $q$-ary $(t,s)$-burst correcting codes.
Section \ref{Sec:Con} establishes the connection between $q$-ary $(t,t)$-burst correcting codes and $q^{t-1}$-ary $(2,2)$-burst correcting codes, and the connection between $q$-ary $(t,s)$-burst correcting codes and $q^{t-s}$-ary $(t',t'-1)$-burst correcting codes where $t > s$ and $t'= \lceil t/(t-s) \rceil+1$, which enables us to only consider the two special types of parameters $s=t-1$ and $s=t$.
Sections \ref{Sec:ttburst} and \ref{Sec:t,t-1burst} present the construction of $(t,t)$-burst correcting codes and $(t,s)$-burst correcting codes, where $t>s$, respectively.
Section \ref{Sec:App} considers several applications of $(t,s)$-burst correcting codes.
Finally Section \ref{Sec:Concl} concludes the paper.

\section{Preliminaries}\label{Sec:Pre}

In this section we introduce some necessary notations which will be useful in this paper.

Unless otherwise stated, we assume $q\geq 2$, $t\geq 0$, and $s\geq 0$ are constant integers.
Let $\Sigma_q^n$ denote the set of $q$-ary sequences of length $n$, where $\Sigma_q=\{0,1,\ldots,q-1\}$.
We use bold letters to represent the sequences and plain letters to represent the symbols in the sequences.
A sequence $\boldsymbol{x}\in\Sigma_q^n$ is denoted either as $x_1x_2x_3\cdots x_n$ or $(x_1,x_2,x_3,\cdots,x_n)$.
Moreover, we set $x_i= 0$ when $i<0$ or $i>n$.
Let $\mathrm{Sum}(\boldsymbol{x})= \sum_{i=1}^n x_i$ and $\mathrm{VT}(\boldsymbol{x})= \sum_{i=1}^{n} i x_i$.
For two integers $i\leq j$, let $[i,j]$ denote the set $\{i,i+1,\ldots,j\}$.
For a set of indices $\mathcal{I}=\{i_1,i_2,\ldots,i_t\}$ with $1 \leq i_1 < i_2 < \cdots < i_t \leq n$, let $\boldsymbol{x}_{\mathcal{I}}$ be the \emph{projection} of $\boldsymbol{x}$ on the indices of $\mathcal{I}$, i.e., $\boldsymbol{x}_{\mathcal{I}}= x_{i_1} x_{i_2} \cdots x_{i_t}$, and $\boldsymbol{x}_{\mathcal{I}}$ is called a \emph{subsequence} of $\boldsymbol{x}$.
Furthermore, if $i_{j+1} = i_j + 1$ for $j \in [1,t-1]$, i.e., $\mathcal{I}$ is an interval, then $\boldsymbol{x}_{\mathcal{I}}$ is called a \emph{substring} of $\boldsymbol{x}$.
Let $|\boldsymbol{x}|$ denote the \emph{length} of $\boldsymbol{x}$ when $\boldsymbol{x}$ is a sequence and let $|\mathcal{I}|$ denote the \emph{size} of $\mathcal{I}$ when $\mathcal{I}$ is a set.

For any sequence $\boldsymbol{x} \in \Sigma_q^n$, we define its \emph{$t \times \lceil n/t \rceil$ array representation} $A_t(\boldsymbol{x})$ as (assuming $n= kt+b$ for some integers $k$ and $b\in [1,t]$):
\begin{equation*}
A_t(\boldsymbol{x})=
    \begin{pmatrix}
        x_1 & x_{t+1} & \cdots & x_{(k-1)t+1} & x_{kt+1} \\
        x_2 & x_{t+2} & \cdots & x_{(k-1)t+2} & x_{kt+2} \\
        \vdots & \vdots & \ddots & \vdots & \vdots \\
        x_b & x_{t+b} & \cdots & x_{(k-1)t+b} & x_{kt+b} \\
        \vdots & \vdots & \ddots & \vdots & \\
        x_t & x_{2t}  & \cdots & x_{kt}
    \end{pmatrix},
\end{equation*} 
and denote $A_t(\boldsymbol{x})_i$ as the $i$-th row of $A_t(\boldsymbol{x})$ for $i \in [1, t]$.
Furthermore, let $\mathcal{D}_{t}(\boldsymbol{x}) = (\mathcal{D}_{t}(\boldsymbol{x})^1,\dots,\mathcal{D}_{t}(\boldsymbol{x})^{\lceil n/t\rceil}) \in \Sigma_{q^t}^{\lceil n/t\rceil}$, where $\mathcal{D}_{t}(\boldsymbol{x})^j$ is the \emph{integer representation} of the $j$-th column of $A_t(\boldsymbol{x})$, i.e., $\mathcal{D}_t(\boldsymbol{x})^j= \sum_{i=1}^{t} q^{i-1} x_{(j-1)t+i}$, for $j \in [1,\lceil n/t\rceil-1]$, and $\mathcal{D}_t(\boldsymbol{x})^{\lceil n/t\rceil}= \sum_{i=1}^{b} q^{i-1} x_{kt+i}$.

For any sequence $\boldsymbol{x} \in \Sigma_q^n$ and any non-negative integer $t$, we say that $\boldsymbol{x}$ suffers \emph{a burst of $t$ consecutive deletions} (or \emph{a $t$-burst-deletion}) if exactly $t$ deletions have occurred consecutively from $\boldsymbol{x}$, i.e., the resultant subsequence is $x_1 \cdots x_{i-1} x_{i+t} \cdots x_{n}$ for some $i \in [1,n-t+1]$.
We say that $\boldsymbol{x}$ suffers \emph{a burst of at most $t$ consecutive deletions} (or \emph{a $^{\leq}t$-burst-deletion}) if it suffers a $t'$-burst-deletion for some $t' \leq t$.
The definition of \emph{a burst of $t$ consecutive insertions} (or \emph{a $t$-burst-insertion}) and \emph{a burst of at most $t$ consecutive insertions} (or \emph{a $^{\leq}t$-burst-insertion}) can be defined analogously.

For non-negative integers $t$ and $s$, a \emph{$t$-deletion-$s$-insertion-burst} (or a \emph{$(t,s)$-burst}) error is a $t$-burst-deletion error followed by an $s$-burst-insertion error at the same coordinate. A $(t,s)$-burst at the $i$-th coordinate, $i \in[1, n-t+1]$, will turn $x_1\dots x_n$ into $x_1\dots x_{i-1}y_1\dots y_sx_{i+t}\dots x_n$, where $y_1\dots y_s$ are the inserted $s$ symbols. For $t\geq 1$ and $s\geq 1$, once the leftmost (rightmost, respectively) deleted and inserted symbols are not distinct, the error can be considered as a burst error of a shorter length. For example, $01010$ can be obtained from $011110$ by a $(4,3)$-burst at the second coordinate (deleting $1111$ and inserting $101$). It could also be seen as a $(2,1)$-burst at the third coordinate (deleting $11$ and inserting $0$). Thus, for $t\geq 1$ and $s\geq 1$, we define a \emph{$t$-deletion-$s$-insertion-exact-burst} (or a \emph{$(t,s)$-exact-burst}) error which further requires the leftmost and rightmost inserted symbols to be different from the leftmost and rightmost deleted symbols, respectively. In addition, a $(t,0)$-burst is a $t$-burst-deletion for $t\geq 1$ and a $(0,s)$-burst is an $s$-burst-insertion for $s\geq 1$.

The \emph{$(t,s)$-burst ball} of $\boldsymbol{x}$, denoted as $\mathcal{B}_{t,s}(\boldsymbol{x})$, is the set consisting of all sequences that can be obtained from $\boldsymbol{x}$ after a $(t,s)$-burst.
A code $\mathcal{C}\subseteq \Sigma_q^n$ is called a \emph{$(t,s)$-burst correcting code} if $\mathcal{B}_{t,s}(\boldsymbol{x})\cap \mathcal{B}_{t,s}(\boldsymbol{y})=\varnothing$ for any two distinct sequences $\boldsymbol{x}$ and $\boldsymbol{y}$ in $\mathcal{C}$. Specially, when $s=0$ or $t=0$, $\mathcal{C}$ is called a \emph{$t$-burst-deletion correcting code} or an \emph{$s$-burst-insertion correcting code}, respectively. The definitions of the other types of error-correcting codes can be defined similarly. To evaluate a code $\mathcal{C}\subseteq \Sigma_q^n$, we calculate its \emph{redundancy} $r(\mathcal{C}) \triangleq n \log q - \log |\mathcal{C}|$, where the logarithm base is $2$.

\section{Sphere Packing Bound on the Code Redundancy}\label{Sec:Bound}

The main challenge in deriving a sphere packing bound for a certain code is to calculate the size of the error ball centered at a given sequence. For deletions the problem becomes more difficult since the size of the error ball is usually dependent on the center, such as the $(t,s)$-burst ball when $s=0$ (i.e., the $t$-burst deletion error ball) given in \cite{Levenshtein-70-BD}. A somehow unexpected result in \cite{Lu-22-arxiv-BD} shows that for the binary alphabet, the $(t,s)$-burst ball is of a constant size independent of the center if $s>0$. 
In this section, we generalize the result of \cite{Lu-22-arxiv-BD} (but prove in a simpler way) and determine the size of the $(t,s)$-burst ball with arbitrary alphabet size.

\begin{theorem}
  Assume $s \neq 0$ and $\boldsymbol{x} \in \Sigma_q^n$, we have $|\mathcal{B}_{t,s}(\boldsymbol{x})| = q^{s-1}((q-1)(n-t+1)+1)$.
\end{theorem}

\begin{IEEEproof}
    Let $\mathcal{B}_{t,s}(\boldsymbol{x},i) \triangleq \{\boldsymbol{y} \in \Sigma_q^{n-t+s}: y_i \neq x_i, \boldsymbol{y}_{[1, n-t+s] \setminus [i,i+s-1]}= \boldsymbol{x}_{[1, n] \setminus [i,i+t-1]}\}$ if $i\in [1,n-t+1]$ and $\mathcal{B}_{t,s}(\boldsymbol{x},i) \triangleq \{\boldsymbol{y} \in \Sigma_q^{n-t+s}: \boldsymbol{y}_{[1,n-t+1]}= \boldsymbol{x}_{[1,n-t+1]} \}$ if $i= n-t+2$.
    We claim that $\{\mathcal{B}_{t,s}(\boldsymbol{x},i): i\in [1,n-t+2]\}$ form a partition of $\mathcal{B}_{t,s}(\boldsymbol{x})$. Obviously, we have $\mathcal{B}_{t,s}(\boldsymbol{x},i) \subseteq \mathcal{B}_{t,s}(\boldsymbol{x})$ and $\mathcal{B}_{t,s}(\boldsymbol{x},i) \cap \mathcal{B}_{t,s}(\boldsymbol{x},j)= \varnothing$, for $i,j\in [1,n-t+2]$ with $i \neq j$.
    It remains to show that $\mathcal{B}_{t,s}(\boldsymbol{x}) \subseteq \bigcup_{i=1}^{n-t+2} \mathcal{B}_{t,s}(\boldsymbol{x},i)$.
    For any $\boldsymbol{y} \in \mathcal{B}_{t,s}(\boldsymbol{x})$, let $j$ be the smallest index on which $\boldsymbol{y}_{[1,n-t+1]}$ and $\boldsymbol{x}_{[1,n-t+1]}$ differ. If such an index $j$ exists then $\boldsymbol{y} \in \mathcal{B}_{t,s}(\boldsymbol{x}, j)$, and otherwise $\boldsymbol{y} \in \mathcal{B}_{t,s}(\boldsymbol{x}, n-t+2)$.
    Therefore, $\{\mathcal{B}_{t,s}(\boldsymbol{x},i): i\in [1,n-t+2]\}$ form a partition of $\mathcal{B}_{t,s}(\boldsymbol{x})$.
    
    Now we calculate the size of $\mathcal{B}_{t,s}(\boldsymbol{x},i)$, where $i \in [1, n-t+1]$, by counting the number of choices for $\boldsymbol{y}$ in $\mathcal{B}_{t,s}(\boldsymbol{x},i)$.
    We observe the following:
    \begin{itemize}
        \item $\boldsymbol{y}_{[1, n-t+s] \setminus [i, i+s-1]} = \boldsymbol{x}_{[1, n] \setminus [i, i+t-1]}$;

        \item There are $q-1$ choices for $y_i$ as $y_i\neq x_i$;

        \item There are $q$ choices for each $y_j$, for $j \in [i+1, i+s-1]$, as there are no restrictions on these values.
    \end{itemize}
    Consequently, the total number of choices for $\boldsymbol{y}$ is given by $|\mathcal{B}_{t,s}(\boldsymbol{x},i)|= (q-1)q^{s-1}$.
    Similarly, we may compute $|\mathcal{B}_{t,s}(\boldsymbol{x},n-t+2)|= q^{s-1}$.
    Therefore, we have
    \begin{align*}
        |\mathcal{B}_{t,s}(\boldsymbol{x})|
        &= \sum_{i=1}^{n-t+1} |\mathcal{B}_{t,s}(\boldsymbol{x},i)|+ |\mathcal{B}_{t,s}(\boldsymbol{x},n-t+2)| \\
        &= (n-t+1)(q-1)q^{s-1}+ q^{s-1} \\
        &= q^{s-1}((q-1)(n-t+1)+1).
    \end{align*}
\end{IEEEproof}

Using the size of the error ball, by a sphere-packing argument, we can obtain an upper bound on the code size, or equivalently a lower bound on the redundancy, for any $(t,s)$-burst correcting code with arbitrary alphabet size.

\begin{theorem}\label{thm:bound}
    For any $(t,s)$-burst correcting code $\mathcal{C} \subseteq \Sigma_q^n$ with $s \neq 0$, we have $|\mathcal{C}| \leq \frac{q^{n-t+1}}{(q-1)(n-t+1)+1}$ and $r(\mathcal{C}) \geq \log{((q-1)(n-t+1)+1)} +(t-1) \log q= \log n +O(1)$.
\end{theorem}

\begin{IEEEproof}
    Since $\mathcal{C}$ is a $(t,s)$-burst correcting code, we have
    \begin{align*}
      \left|\bigcup_{\boldsymbol{x} \in \mathcal{C}} \mathcal{B}_{t,s}(\boldsymbol{x})\right|
      &= \sum_{\boldsymbol{x} \in \mathcal{C}} |\mathcal{B}_{t,s}(\boldsymbol{x})|\\
      &= |\mathcal{C}| \cdot q^{s-1}((q-1)(n-t+1)+1).
    \end{align*}
    Observe that $\mathcal{B}_{t,s}(\boldsymbol{x}) \subseteq \Sigma_q^{n-t+s}$ for $\boldsymbol{x}\in \mathcal{C}$, we get 
    \begin{align*}
      \left|\bigcup_{\boldsymbol{x} \in \mathcal{C}} \mathcal{B}_{t,s}(\boldsymbol{x}) \right| \leq q^{n-t+s}.
    \end{align*}
    Therefore, we can compute $|\mathcal{C}| \leq \frac{q^{n-t+s}}{q^{s-1}((q-1)(n-t+1)+1)}$ and $r(\mathcal{C}) \geq \log{((q-1)(n-t+1)+1)} +(t-1) \log q$.
\end{IEEEproof}

\begin{remark}
   When $t=s=1$, the lower bound established in Theorem \ref{thm:bound} indeed corresponds to the sphere packing bound for single-substitution correcting codes. 
   Furthermore, since there is an equivalence between $(t,s)$-burst correcting codes and $(s,t)$-burst correcting codes \cite{Lu-22-arxiv-BD}, our lower bound can be expressed as $\log{((q-1)(n-k+1)+1)} +(k-1) \log q$, where $k=\min\{t,s\}$.
\end{remark}

For $s=0$, Wang \emph{et al.} \cite{Wang-22-arXiv-BD} have shown that the redundancy of any $q$-ary $(t,0)$-burst correcting code is lower bounded by $\log (q-1)(n-2t+1)+ (t-1)\log q- \log (1-\frac{1}{q^{n-2t+1}})$. Thus, combining with our results, we have the following corollary.

\begin{corollary}
  For any $(t,s)$-burst correcting code $\mathcal{C} \subseteq \Sigma_q^n$, we have $r(\mathcal{C}) \geq \log n +O(1)$.
\end{corollary}

\begin{remark}
   The lower bound $\log n+O(1)$ can be anticipated from the existential results, as any $(t,s)$-burst correcting code is a $(t-s,0)$-burst correcting code when $t>s$ and a single-substitution correcting code when $t=s$. 
   Each of these two types of codes is lower bounded by $\log n+O(1)$.
   It is clear that when $t\geq s$, the bound established in Theorem \ref{thm:bound} is superior to simply applying the existential results directly.
 \end{remark}
 
In the rest of the paper our objective is to construct $(t,s)$-burst correcting codes that asymptotically achieve the aforementioned lower bound.
Since it has been shown in \cite{Lu-22-arxiv-BD} that any $(t,s)$-burst correcting code is also an $(s,t)$-burst correcting code, we assume $t \geq s$.
Moreover, we assume $t\geq 2$ since the $(1,1)$-burst is simply one substitution and the $(1,0)$-burst is simply one deletion, and both of them have been well understood.

\section{Connections Between Several Types of Burst Errors}\label{Sec:Con}

In this section, we establish the connection between $q$-ary $(t,t)$-burst correcting codes and $q^{t-1}$-ary $(2,2)$-burst correcting codes, as well as the connection between $q$-ary $(t,s)$-burst correcting codes and $q^{t-s}$-ary $(t',t'-1)$-burst correcting codes where $t > s$ and $t'= \lceil t/(t-s) \rceil+1$.
Once these two connections are established,
it suffices to consider only two types of errors, the $(2,2)$-burst and the $(t,t-1)$-burst. Firstly, we give the following simple but useful observation.

\begin{observation}\label{obs:ET}
For any sequence $\boldsymbol x$, an interval of length $t \geq 2$ spans at most $\lceil t/(t-s) \rceil+1$ adjacent columns in the array $A_{t-s}(\boldsymbol x)$ when $t>s$, and exactly two adjacent columns in the array $A_{t-1}(\boldsymbol x)$.
\end{observation}

\subsection{$(t,t)$-Burst and $(2,2)$-Burst where $t \geq 2$}

Assume $t \geq 2$. For any $\boldsymbol{x} \in \Sigma_q^n$, we consider its $(t-1) \times \lceil n/(t-1) \rceil$ array representation $A_{t-1}(\boldsymbol{x})$ and let $\mathcal{D}_{t-1}(\boldsymbol{x}) \in \Sigma_{q^{t-1}}^{\lceil n/(t-1) \rceil}$, where $\mathcal{D}_{t-1}(\boldsymbol{x})^j$ is the integer representation of the $j$-th column of $A_{t-1}(\boldsymbol{x})$. Now, we determine the type of error that occurs in $\mathcal{D}_{t-1}(\boldsymbol{x})$ when $\boldsymbol{x}$ suffers a $(t,t)$-burst.

\begin{lemma}\label{lem:ET_tt}
Assume $t \geq 2$. $\mathcal{D}_{t-1}(\boldsymbol{x})$ suffers a $(2,2)$-burst if $\boldsymbol{x}$ suffers a $(t,t)$-burst.
\end{lemma}

\begin{IEEEproof}
If $\boldsymbol{x}$ suffers a $(t,t)$-burst error, by Observation \ref{obs:ET}, in $A_{t-1}(\boldsymbol{x})$ the total number of columns remains unchanged and
at most two adjacent columns are affected. Thus, $\mathcal{D}_{t-1}(\boldsymbol{x})$ suffers a $(2,2)$-burst.
\end{IEEEproof}

Immediately we have the following theorem.
\begin{theorem}\label{thm:ET_tt}
Assume $t \geq 2$. $\mathcal{C} \subseteq \Sigma_q^n$ is a $q$-ary $(t,t)$-burst correcting code if $\mathcal{D}_{t-1}(\mathcal{C}) \triangleq \{ \mathcal{D}_{t-1}(\boldsymbol{x}): \boldsymbol{x} \in \mathcal{C} \}$ is a $q^{t-1}$-ary $(2,2)$-burst correcting code.
\end{theorem}

\begin{IEEEproof}
    Suppose $\boldsymbol{x} \in \mathcal{C}$ suffers a $(t,t)$-burst error and results in $\boldsymbol{z}$. By Lemma \ref{lem:ET_tt}, we have $\mathcal{D}_{t-1}(\boldsymbol{z}) \in \mathcal{B}_{2,2}(\mathcal{D}_{t-1}(\boldsymbol{x}))$.
    Since $\mathcal{D}_{t-1}(\mathcal{C})$ is a $(2,2)$-burst-correcting code, we can correctly recover $\mathcal{D}_{t-1}(\boldsymbol{x})$ from $\mathcal{D}_{t-1}(\boldsymbol{z})$.
    Then we can recover the original sequence $\boldsymbol{x}$ from the corrected $\mathcal{D}_{t-1}(\boldsymbol{x})$, as the map from $\boldsymbol{x} \in \Sigma_q^n$ to $\mathcal{D}_{t-1}(\boldsymbol{x}) \in \Sigma_{q^{t-1}}^{\lceil n/(t-1) \rceil}$ is injective.
    Therefore, $\mathcal{C}$ is a $(t,t)$-burst correcting code.
\end{IEEEproof}

\subsection{$(t,s)$-Burst and $(t',t'-1)$-Burst where $t>s$ and $t'= \lceil t/(t-s) \rceil+1$}

For any $\boldsymbol{x} \in \Sigma_q^n$, we consider its $(t-s) \times \lceil n/(t-s) \rceil$ array representation $A_{t-s}(\boldsymbol{x})$ and let $\mathcal{D}_{t-s}(\boldsymbol{x}) \in \Sigma_{q^{t-s}}^{\lceil n/(t-s) \rceil}$.
Now, we determine the type of error that occurs in $\mathcal{D}_{t-s}(\boldsymbol{x})$ when $\boldsymbol{x}$ suffers a $(t,s)$-burst.

\begin{lemma}\label{lem:ET_ts}
Assume $t>s$ and $t'= \lceil t/(t-s) \rceil+1$. $\mathcal{D}_{t-s}(\boldsymbol{x})$ suffers a $(t',t'-1)$-burst if $\boldsymbol{x}$ suffers a $(t,s)$-burst.
\end{lemma}

\begin{IEEEproof}
    If $\boldsymbol{x}$ suffers a $(t,s)$-burst, by Observation \ref{obs:ET}, in $A_{t-s}(\boldsymbol{x})$ the total number of columns decreases by one and
at most $t'$ adjacent columns are affected. Thus, $\mathcal{D}_{t-s}(\boldsymbol{x})$ suffers a $(t',t'-1)$-burst.
\end{IEEEproof}

Immediately we have the following theorem.
\begin{theorem}\label{thm:ET_ts}
Assume $t>s$ and $t'= \lceil t/(t-s) \rceil +1$. $\mathcal{C} \subseteq \Sigma_q^n$ is a $q$-ary $(t,s)$-burst correcting code if $\mathcal{D}_{t-s}(\mathcal{C}) \triangleq \{ \mathcal{D}_{t-s}(\boldsymbol{x}): \boldsymbol{x} \in \mathcal{C} \}$ is a $q^{t-s}$-ary $(t',t'-1)$-burst correcting code.
\end{theorem}

\begin{IEEEproof}
    Suppose $\boldsymbol{x} \in \mathcal{C}$ suffers a $(t,s)$-burst error and results in $\boldsymbol{z}$. Then by Lemma \ref{lem:ET_ts}, we have $\mathcal{D}_{t-s}(\boldsymbol{z}) \in \mathcal{B}_{t',t'-1}(\mathcal{D}_{t-s}(\boldsymbol{x}))$.
    Since $\mathcal{D}_{t-s}(\mathcal{C})$ is a $(t',t'-1)$-burst-correcting code, we can correctly recover $\mathcal{D}_{t-s}(\boldsymbol{x})$ from $\mathcal{D}_{t-s}(\boldsymbol{z})$.
    Subsequently, we can recover the original sequence $\boldsymbol{x}$ from the corrected $\mathcal{D}_{t-s}(\boldsymbol{x})$, as the map from $\boldsymbol{x} \in \Sigma_q^n$ to $\mathcal{D}_{t-s}(\boldsymbol{x}) \in \Sigma_{q^{t-s}}^{\lceil n/(t-s) \rceil}$ is injective.
    Therefore, $\mathcal{C}$ is a $(t,s)$-burst correcting code.
\end{IEEEproof}

\begin{remark}\label{rmk:connection}
By Theorems \ref{thm:ET_tt} and \ref{thm:ET_ts}, we will focus on the constructions of $(2,2)$-burst correcting codes and $(t,t-1)$-burst correcting codes, which will give rise to the constructions of general $(t,t)$-burst correcting codes and $(t,s)$-burst correcting codes.
\end{remark}

\section{$(t,t)$-Burst Correcting Codes}\label{Sec:ttburst}

It is well-known that Fire codes \cite{Fire-59-BS,Blahut} are binary codes that can correct a $(t,t)$-burst error.
In this section, our aim is not to generalize them to a non-binary alphabet, but rather to seek a new method for constructing $q$-ary $(t,t)$-burst correcting codes with a simple decoding algorithm.
We believe that this new method may have its own potential in constructing other types of burst-error correcting codes.

Assume $(t-1)|n$, and otherwise we append zeros at the very end of each sequence such that its length is the smallest integer greater than $n$ and divisible by $(t-1)$.
By Theorem \ref{thm:ET_tt}, we first construct $(2,2)$-burst correcting codes for arbitrary alphabet size $q$, and then use them to give rise to $(t,t)$-burst correcting codes for $t \geq 2$.

Here, we view any sequence $\boldsymbol{x}$ as a $2 \times \lceil n/2 \rceil$ array $A_{2}(\boldsymbol{x})$.
Obviously, a $(2,2)$-burst in $\boldsymbol{x}$ results in at most one substitution in each row of $A_{2}(\boldsymbol{x})$.
A simple method to combat such an error is to encode each row of $A_{2}(\boldsymbol{x})$ as a $q$-ary single substitution correcting code, independently.
However, this demands $2 \log n +O(1)$ bits of redundancy and does not take advantage of the consecutiveness of the two erroneous coordinates.
Below we propose a novel approach to combat a $(2,2)$-burst error, which only incurs $\log n +O(1)$ bits of redundancy: a sum constraint is imposed on each row separately, whereas a VT type constraint is imposed on the two rows together instead of on each individual row.

\begin{construction}\label{con:(2,2)-burst}
  Let $\boldsymbol{a} = (a_1,a_2,a_3)$, where $a_1,a_2\in \mathbb{Z}_{2q}, a_3\in \mathbb{Z}_{q(q-1)(n-1)+1}$, we define the code $\mathcal{C}_{2,2}(n,q;\boldsymbol{a})$ as
  \begin{equation*}
  \begin{aligned}
    \Big\{\boldsymbol{x} \in \Sigma_q^n: &~ \mathrm{Sum}(A_2(\boldsymbol{x})_1) \equiv a_1 \pmod{2q}, \\
    &~ \mathrm{Sum}(A_2(\boldsymbol{x})_2) \equiv a_2 \pmod{2q}, \\
    &~ \mathrm{WVT}(\boldsymbol{x}) \equiv a_3 \pmod{q(q-1)(n-1)+1} \Big\},
  \end{aligned}
  \end{equation*}
  where $\mathrm{WVT}(\boldsymbol{x})= \mathrm{VT}(A_2(\boldsymbol{x})_1)+ (2q-1) \mathrm{VT}(A_2(\boldsymbol{x})_2)$.
\end{construction}

\begin{theorem}\label{thm:(2,2)-burst}
  The code $\mathcal{C}_{2,2}(n,q;\boldsymbol{a})$ is a $q$-ary $(2,2)$-burst correcting code. Moreover, there exists a choice of parameters, such that the code $\mathcal{C}_{2,2}(n,q;\boldsymbol{a})$ has at most $\log n + 4 \log q + 2$ bits of redundancy.
\end{theorem}

\begin{IEEEproof}
    We prove that $\mathcal{C}_{2,2}(n,q;\boldsymbol{a})$ is a $q$-ary $(2,2)$-burst correcting code by providing a decoding algorithm.
    Suppose $\boldsymbol{x}\in \mathcal{C}_{2,2}(n,q;\boldsymbol{a})$ suffers a $(2,2)$-burst error at the $i$-th coordinate, where $i\in [1,n-1]$, and results in $\boldsymbol{z}$. 
    We consider the $2 \times \lceil n/2 \rceil$ array representation of $\boldsymbol{x}$.
    Let $\delta_k\in [1-q,q-1]$, for $k\in \{1,2\}$, be the magnitude of error in the $k$-th row of $A_{2}(\boldsymbol{x})$. 
    We have $\delta_k\equiv \mathrm{Sum}(A_2(\boldsymbol{z})_k) -a_k \pmod{2q}$.
    To recover $\boldsymbol{x}$ from $\boldsymbol{z}$, it remains to find $i$. 
    Let $\Delta\triangleq \mathrm{WVT}(\boldsymbol{z})- \mathrm{WVT}(\boldsymbol{x})= \delta_1 \left\lceil \frac{i+1}{2} \right\rceil+ (2q-1)\delta_2 \left\lfloor \frac{i+1}{2} \right\rfloor$. We can determine $\Delta$ from $a_3$ and $\boldsymbol{z}$ since
    \begin{itemize}
      \item The sign of $\Delta$ can be determined (if $\delta_2>0$, or if $\delta_2=0$ and $\delta_1\geq 0$, then $\Delta\geq 0$; 
      if $\delta_2<0$, or if $\delta_2=0$ and $\delta_1<0$, then $\Delta<0$);
      \item $\mathrm{WVT}(\boldsymbol{z})$ is at most $(q-1)\lceil n/2 \rceil + (2q-1)(q-1)\lfloor n/2 \rfloor \leq q(q-1)(n-1)$ away from $\mathrm{WVT}(\boldsymbol{x})$.
    \end{itemize}
    Then there are at most two possible choices for $i$:
    \begin{itemize}
        \item $i= \frac{2\Delta}{\delta_1+(2q-1)\delta_2}-1$ if $2\nmid i$; 

        \item $i= \frac{2(\Delta-\delta_1)}{\delta_1+(2q-1)\delta_2}$ if $2|i$. 
    \end{itemize}
    When $\delta_2=0$, it means that there is no error in the second row of $A_{2}(\boldsymbol{x})$, and it suffices to identify the erroneous coordinate in the first row of $A_{2}(\boldsymbol{x})$; thus, the two cases are essentially the same. 
    When $\delta_2\neq 0$, since $i$ is an integer, only one of the two cases can hold, thus we can identify it uniquely.
    Then we can recover $\boldsymbol{x}$.

    Note that $\bigcup_{\boldsymbol{a} \in \Sigma_{2q}^2 \times \Sigma_{q(q-1)(n-1)+1}} \mathcal{C}_{2,2}(n,q;\boldsymbol{a})= \Sigma_q^n$, then by the pigeonhole principle, we can choose some $\boldsymbol{a}$ such that the size of $\mathcal{C}_{2,2}(n,q;\boldsymbol{a})$ is at least $\frac{q^{n}}{(2q)^2\cdot (q(q-1)(n-1)+1)}$ and the corresponding redundancy is at most $\log n + 4 \log q + 2$.
\end{IEEEproof}

By Theorem \ref{thm:ET_tt}, we can utilize a $q^{t-1}$-ary $(2,2)$-burst correcting code of length $\frac{n}{t-1}$ to construct a $q$-ary $(t,t)$-burst correcting code of length $n$.
Combining this with Theorem \ref{thm:(2,2)-burst}, we have the following conclusion.

\begin{theorem}\label{thm:(t,t)-burst}
  Assume $(t-1)|n$. Let $\mathcal{C}_{2,2}(\frac{n}{t-1},q^{t-1};\boldsymbol{a})$ be the $q^{t-1}$-ary $(2,2)$-burst correcting code constructed in Construction \ref{con:(2,2)-burst}, where $\boldsymbol{a} \in \Sigma_{2 q^{t-1}}^2 \times \Sigma_{q^{t-1}(q^{t-1}-1)(\frac{n}{t-1}-1)+1}$, then the code
  \begin{equation*}
  \begin{aligned}
    \mathcal{C}_{t,t}(n,q;\boldsymbol{a})= \Big\{\boldsymbol{x} \in \Sigma_q^n: \mathcal{D}_{t-1}(\boldsymbol{x}) \in \mathcal{C}_{2,2}(\tfrac{n}{t-1},q^{t-1};\boldsymbol{a}) \Big\},
  \end{aligned}
  \end{equation*}
  is a $q$-ary $(t,t)$-burst correcting code and there exists a choice of $\boldsymbol{a}$ such that the corresponding redundancy is at most $\log \frac{n}{t-1}+ 4 \log q^{t-1}+ 2 = \log n + 4(t-1) \log q + 2 -\log (t-1)$.
\end{theorem}

\section{$(t,s)$-Burst Correcting Codes}\label{Sec:t,t-1burst}

In this section we consider $(t,s)$-burst correcting codes for the case $t>s$. Following Theorem \ref{thm:ET_ts}, it suffices to consider $s=t-1$.

\subsection{Binary $(t,t-1)$-Burst Correcting Codes}

We first consider the binary alphabet.
A $(t,t-1)$-burst error can be seen as a single deletion followed by a $(t-1)$-burst-substitution.
Thus, to combat a $(t,t-1)$-burst error, a natural idea is to start from some well-designed single deletion correcting code and then impose some additional constraints.
When $t=2$, Schoeny \emph{et al.} \cite{Schoeny-17-IT-BD} demonstrated that a straightforward sum constraint modulo four on the sequences themselves is sufficient for this purpose. 
However, for $t\geq 3$, the problem becomes considerably more complex, and no efficient solutions have been identified to date.
In the following, we will present three distinct types of sum constraints that, when employed collectively, can facilitate the construction of binary $(t,t-1)$-burst correcting codes.

\subsubsection{The First Constraint}

Use a sum constraint on the transmitted sequence itself to determine the number of ones in it when it suffers a $(t,t-1)$-burst.

\begin{lemma}\label{lem:variation1}
Let $\boldsymbol{x} \in \Sigma_2^n$ satisfy the sum constraint $\mathrm{Sum}(\boldsymbol{x}) \equiv a_0 \pmod{2t}$ where $a_0 \in \mathbb{Z}_{2t}$.
When $\boldsymbol{x}$ suffers a $(t,t-1)$-burst and results in $\boldsymbol{x}'$, we can determine the number of ones in $\boldsymbol{x}$ as $\mathrm{Sum}(\boldsymbol{x}')-\Delta$, where $\Delta\in[-t,t-1]$ and $\Delta\equiv\mathrm{Sum}(\boldsymbol{x}')-a_0 \pmod{2t}$.
\end{lemma}

\begin{IEEEproof}
When $\boldsymbol{x}$ suffers a $(t,t-1)$-burst, the number of ones increases by at most $t-1$ and decreases by at most $t$.
The decoder observes $\mathrm{Sum}(\boldsymbol{x}')$ and determines the change in the number of occurrences of ones as $\Delta\in[-t,t-1]$, where $\Delta\equiv\mathrm{Sum}(\boldsymbol{x}')-a_0 \pmod{2t}$. Then the number of ones in $\boldsymbol{x}$ is exactly $\mathrm{Sum}(\boldsymbol{x}')-\Delta$.
\end{IEEEproof}

\begin{remark}
In Lemma \ref{lem:variation1}, the number of ones in $\boldsymbol{x}$ is determined by $\boldsymbol{x}'$ and $a_0$.
That is to say, if $\mathrm{Sum}(\boldsymbol{x}) \equiv \mathrm{Sum}(\boldsymbol{y}) \pmod{2t}$ and $\mathcal{B}_{t,t-1}(\boldsymbol{x}) \cap \mathcal{B}_{t,t-1}(\boldsymbol{y}) \neq \varnothing$,
then $\boldsymbol{x}$ and $\boldsymbol{y}$ must have the same number of ones.
\end{remark}

\subsubsection{The Second Constraint}

Use a sum constraint on each row of the array representation to guarantee that the distance between the leftmost and rightmost indices on which two codewords differ is large.

\begin{lemma}\label{lem:dist_1}
Let $\boldsymbol{x}$ and $\boldsymbol{y}$ be distinct sequences of length $n$ with non-negative integer entries and $|x_i- y_i|<q$ for $i\in [1,n]$.
Given a positive integer $k$,
if $\mathrm{Sum}(A_k(\boldsymbol{x})_j) \equiv \mathrm{Sum}(A_k(\boldsymbol{y})_j) \pmod{q}$ for $j\in [1,k]$,
then the distance between the leftmost and rightmost indices on which $\boldsymbol{x}$ and $\boldsymbol{y}$ differ is at least $k$.
\end{lemma}

\begin{IEEEproof}
Since $\boldsymbol{x} \neq \boldsymbol{y}$, we can find the leftmost index $\ell_1$ on which $\boldsymbol{x}$ and $\boldsymbol{y}$ differ.
Due to the sum constraint on each row, and the premise that $|x_i-y_i|<q$ for $i\in [1,n]$, there must exist another index $\ell_2$ with $\ell_2\equiv \ell_1 \pmod{k}$ on which $\boldsymbol{x}$ and $\boldsymbol{y}$ differ. Then $\ell_2- \ell_1\geq k$, which completes the proof.
\end{IEEEproof}

\begin{remark}\label{rmk:dist}
It should be noted that Lemma \ref{lem:dist_1} also holds when we replace the premise that $|x_i-y_i|<q$ for $i\in [1,n]$ simply as $\boldsymbol{x}, \boldsymbol{y} \in \Sigma_q^n$. We present the lemma for a general version since it will be used in Lemma \ref{lem:dist_2}.
\end{remark}

Before introducing the third constraint, we define the following marker sequence, which records the coordinates of all the ones in a sequence.

\begin{definition}
For any sequence $\boldsymbol{x} \in \Sigma_2^n$, assume the number of ones in $\boldsymbol{x}$ is $m$, i.e., $\mathrm{Sum}(\boldsymbol{x})= m$. Define the following marker sequence
\begin{equation*}
    \mathcal{L}^{\boldsymbol{x}}= \mathcal{L}_1^{x} \mathcal{L}_2^{x} \cdots \mathcal{L}_m^{x},
\end{equation*}
where $\mathcal{L}_i^{x}$ denotes the coordinate of the $i$-th one in $\boldsymbol{x}$, for $i\in [1,m]$.
\end{definition}

\begin{example}
  Let $\boldsymbol{x}= 01010011 \in \Sigma_2^8$, then $\mathcal{L}^{\boldsymbol{x}}= 2478$.
\end{example}

\subsubsection{The Third Constraint}
Use a sum constraint on the array representation of the marker sequences for two codewords $\boldsymbol{x}$ and $\boldsymbol{y}$ of the same weight, to guarantee that the distance between the leftmost and rightmost indices on which $\mathcal{L}^{\boldsymbol{x}}$ and $\mathcal{L}^{\boldsymbol{y}}$ differ is large.

\begin{lemma}\label{lem:dist_2}
    For two distinct sequences $\boldsymbol{x}$ and $\boldsymbol{y}$ in $\Sigma_2^n$ and two positive integers $t$ and $k$, suppose it holds that
    \begin{itemize}
      \item $m \triangleq \mathrm{Sum}(\boldsymbol{x}) = \mathrm{Sum}(\boldsymbol{y})$,
      \item $|\mathcal{L}_j^{\boldsymbol{x}}- \mathcal{L}_j^{\boldsymbol{y}}| < t$ for $j \in [1,m] $,
      \item $\mathrm{Sum}(A_k(\mathcal{L}^{\boldsymbol{x}})_j) \equiv \mathrm{Sum}(A_k(\mathcal{L}^{\boldsymbol{y}})_j) \pmod{t}$ for $j\in [1, k]$,
    \end{itemize}
    then the distance between the leftmost and rightmost indices on which $\mathcal{L}^{\boldsymbol{x}}$ and $\mathcal{L}^{\boldsymbol{y}}$ differ is at least $k$.
\end{lemma}

\begin{IEEEproof}
Observe that $\mathcal{L}^{\boldsymbol{x}}$ and $\mathcal{L}^{\boldsymbol{y}}$ are sequences of length $m$ with non-negative integer entries, and $|\mathcal{L}_j^{\boldsymbol{x}}-\mathcal{L}_j^{\boldsymbol{y}}| < t$ for $j\in [1,m]$. The conclusion follows directly from Lemma \ref{lem:dist_1}.
\end{IEEEproof}

Now, we use these three constraints above to construct binary $(t,t-1)$-burst correcting codes.
As with most literatures concerning deletions and insertions, we consider a more general version of $P$-bounded codes.

\begin{definition}
Let $P\leq n$. $\mathcal{C} \subseteq \Sigma_q^n$ is called a $P$-bounded $(t,s)$-burst correcting code if, for any two distinct sequences $\boldsymbol{x}=\boldsymbol{u} \tilde{\boldsymbol{x}}\boldsymbol{v},  \boldsymbol{y}=\boldsymbol{u} \tilde{\boldsymbol{y}}\boldsymbol{v} \in \mathcal{C}$ with $|\tilde{\boldsymbol{x}}|= |\tilde{\boldsymbol{y}}|\leq P$, it holds that $\mathcal{B}_{t,s}(\tilde{\boldsymbol{x}}) \cap \mathcal{B}_{t,s}(\tilde{\boldsymbol{y}})= \varnothing$. 
In other words, $\mathcal{C} \subseteq \Sigma_q^n$ is called a $P$-bounded $(t, s)$-burst correcting code if it can correct a $(t, s)$-burst given the additional knowledge of an interval of length $P$ containing all the erroneous coordinates. 
\end{definition}

\begin{remark}
    When $P=n$, $P$-bounded $(t,s)$-burst correcting codes are essentially $(t,s)$-burst correcting codes.
\end{remark}

\begin{construction}\label{con:(t,t-1)-burst}
For any positive integer $t$, set $k= \lfloor \frac{t^2}{2} \rfloor$. Assume $P$ and $n$ are sufficiently large and $P\leq n$. Let $\boldsymbol{a} \in \mathbb{Z}_{tP} \times \mathbb{Z}_{2t}$, $\boldsymbol{b} \in \mathbb{Z}_2^{k}$, and $\boldsymbol{c}, \boldsymbol{c}' \in \mathbb{Z}_{t}^k$. We define the code $\mathcal{C}_{t,t-1}(n,P,2;\boldsymbol{a},\boldsymbol{b},\boldsymbol{c},\boldsymbol{c}')$ as
  \begin{equation*}
  \begin{aligned}
  \Big\{\boldsymbol{x} \in \Sigma_2^n:
    &~ \mathrm{VT}(\boldsymbol{x}) \equiv a_1 \pmod{tP}, \\
    &~ \mathrm{Sum}(\boldsymbol{x}) \equiv a_2 \pmod{2t}, \\
    &~ \mathrm{Sum}(A_{k}(\boldsymbol{x})_j) \equiv b_j \pmod{2}, ~\forall j\in [1,k], \\
    &~ \mathrm{Sum}(A_k(\mathcal{L}^{\boldsymbol{x}})_j) \equiv c_j \pmod{t}, ~\forall j\in [1,k], \\
    &~ \mathrm{Sum}(A_k( \mathcal{L}^{\overline{\boldsymbol{x}}} )_j) \equiv c_j' \pmod{t}, ~\forall j\in [1,k] \Big\},
  \end{aligned}
  \end{equation*}
where $\overline{\boldsymbol{x}} \triangleq (1-x_1,1-x_2,\ldots,1-x_n)$.
\end{construction}

\begin{theorem}\label{thm:(t,t-1)-burst}
  The code $\mathcal{C}_{t,t-1}(n,P,2;\boldsymbol{a},\boldsymbol{b},\boldsymbol{c},\boldsymbol{c}')$ is a binary $P$-bounded $(t,t-1)$-burst correcting code. Moreover, there exists a choice of parameters, such that the code $\mathcal{C}_{t,t-1}(n,P,2;\boldsymbol{a},\boldsymbol{b},\boldsymbol{c},\boldsymbol{c}')$ has at most $\log P + O(1)$ bits of redundancy.
\end{theorem}

\begin{IEEEproof}
We prove it by contradiction. 
Assume there exist two distinct codewords $\boldsymbol{x}$ and $\boldsymbol{y}$ in $\mathcal{C}_{t,t-1}(n,P,2;\boldsymbol{a},\boldsymbol{b},\boldsymbol{c},\boldsymbol{c}')$, such that $\mathcal{B}_{t,t-1}(\boldsymbol{x}) \cap \mathcal{B}_{t,t-1}(\boldsymbol{y}) \neq \varnothing$. For any $\boldsymbol{z} \in \mathcal{B}_{t,t-1}(\boldsymbol{x}) \cap \mathcal{B}_{t,t-1}(\boldsymbol{y})$, there exist two indices $i$ and $j$ (without loss of generality assume $i \leq j$), such that
\begin{equation}\label{eq:z}
\begin{aligned}
\boldsymbol{z} 
&= x_1 \cdots x_{i-1} \underline{x_i' \cdots x_{i+t-2}'} x_{i+t} \cdots x_{j+t-1} x_{j+t} \cdots x_n \\
&= y_1 \cdots y_{i-1} y_i \cdots y_{j-1} \underline{y_j' \cdots y_{j+t-2}'} y_{j+t} \cdots y_n.
\end{aligned}
\end{equation}
It holds that $x_{\ell} = y_{\ell}$ for $\ell \in [1,i-1] \cup [j+t,n]$.
By Lemma \ref{lem:dist_1} and the sum constraint on each row of $A_k(\boldsymbol{x})$ and $A_k(\boldsymbol{y})$, we obtain $j+t-1-i \geq k = \lfloor \frac{t^2}{2} \rfloor \geq 2t-2$, i.e., $j+1 \geq i+t$. Thus, Figure \ref{Fig:1} holds and we have $(x_i' \cdots x_{i+t-2}')=(y_i \cdots y_{i+t-2})$ and $(y_j' \cdots y_{j+t-2}')=(x_{j+1} \cdots x_{j+t-1})$.
Let $\Delta \in [-t,t-1]$ be such that $\Delta\equiv \mathrm{Sum}(\boldsymbol{z})-a_2 \pmod{2t}$, we have the following two expressions of $\Delta$.
\begin{itemize}
    \item Since $\boldsymbol{z} \in \mathcal{B}_{t,t-1}(\boldsymbol{x})$, by Lemma \ref{lem:variation1}, we have $\Delta= \mathrm{Sum}(\boldsymbol{z})- \mathrm{Sum}(\boldsymbol{x})= \sum_{\ell=i}^{i+t-2}y_{\ell}- \sum_{\ell=i}^{i+t-1} x_{\ell}$.

    \item Since $\boldsymbol{z} \in \mathcal{B}_{t,t-1}(\boldsymbol{y})$, by Lemma \ref{lem:variation1}, we have $\Delta= \mathrm{Sum}(\boldsymbol{z})- \mathrm{Sum}(\boldsymbol{y})= \sum_{\ell=j+1}^{j+t-1}x_{\ell}- \sum_{\ell=j}^{j+t-1} y_{\ell}$.
\end{itemize}
Now we calculate $\mathrm{VT}(\boldsymbol{x})- \mathrm{VT}(\boldsymbol{y})$.
Let $A= \sum_{\ell=i}^{i+t-1} \ell x_{\ell}- \sum_{\ell=i}^{i+t-2} \ell y_{\ell}$, $B= \sum_{\ell=i+t}^{j} x_{\ell}$, and $C= \sum_{\ell=j+1}^{j+t-1} \ell x_{\ell}- \sum_{\ell=j}^{j+t-1} \ell y_{\ell}$.
According to Figure \ref{Fig:1}, we have $\mathrm{VT}(\boldsymbol{x})- \mathrm{VT}(\boldsymbol{y})= A+B+C$.
Note that the additional knowledge of $P$-boundness indicates that $j+t-i\leq P$, then $0 \leq B\leq j-i-t+1 \leq P-2t+1$.

\begin{figure*}[htbp]
  \centering
    \begin{tikzpicture}
        \matrix (A) [matrix of math nodes,row sep=.5cm]
        {
        x_1 & \cdots & x_{i-1} & x_{i} & \cdots & x_{i+t-2} & x_{i+t-1} & x_{i+t} & \cdots & x_{j} & x_{j+1} & \cdots & x_{j+t-1} & x_{j+t} & \cdots & x_n\\
        y_1 & \cdots & y_{i-1} & y_{i} & \cdots & y_{i+t-2} & y_{i+t-1} & \cdots & y_{j-1} & y_{j} & y_{j+1} & \cdots & y_{j+t-1} & y_{j+t} & \cdots & y_n\\
        };
        \draw [black,thick] (A-1-1) -- (A-2-1);
        \draw [black,thick] (A-1-3) -- (A-2-3);
        \draw [black,thick]  (A-1-8) -- (A-2-7);
        \draw [black,thick]  (A-1-10) -- (A-2-9);
        \draw [black,thick]  (A-1-14) -- (A-2-14);
        \draw [black,thick] (A-1-16) -- (A-2-16);
    \end{tikzpicture}
    \caption{Illustration of sequences $\boldsymbol{x}$ and $\boldsymbol{y}$ satisfying Equation (\ref{eq:z}) and the condition $j+1\geq i+t$, where an edge indicates that the two symbols are identical.}
    \label{Fig:1}
\end{figure*}

The contradiction is found by showing that $A+B+C \equiv 0 \pmod{tP}$ does not hold,
which contradicts to the VT constraint of the code. For this purpose we proceed with the following four cases, depending on the value of $\Delta$.

\medskip

\begin{itemize}
  \item $\Delta= 0$, that is, $\boldsymbol{x}_{[i,i+t-1]}$ and $\boldsymbol{y}_{[i,i+t-2]}$ have the same number of ones, and the same holds for $\boldsymbol{x}_{[j+1,j+t-1]}$ and $\boldsymbol{y}_{[j,j+t-1]}$. Thus, $\boldsymbol{x}$ and $\boldsymbol{y}$ have the same number of ones. Let $\mathcal{L}^{x}_{\ell}$ and $\mathcal{L}^{y}_{\ell}$ be the coordinates of the $\ell$-th one in $\boldsymbol{x}$ and $\boldsymbol{y}$, respectively, then we have
          \begin{equation*}
            \begin{cases}
              \mathcal{L}^{x}_{\ell}= \mathcal{L}^{y}_{\ell}, & \mbox{if } \mathcal{L}^{x}_{\ell} \in [1,i-1] \cup [j+t,n]; \\
              \mathcal{L}^{x}_{\ell}= \mathcal{L}^{y}_{\ell}+1, & \mbox{if } \mathcal{L}^{x}_{\ell} \in [i+t,j]; \\
              |\mathcal{L}^{x}_{\ell} - \mathcal{L}^{y}_{\ell}| \leq t-1, & \mbox{otherwise},
            \end{cases}
          \end{equation*}
          which indicates that the maximum of the absolute values of the entries in $\mathcal{L}^{\boldsymbol{x}}- \mathcal{L}^{\boldsymbol{y}}$ is at most $t-1$.
          Then by Lemma \ref{lem:dist_2}, the distance between the leftmost and rightmost indices on which $\mathcal{L}^{\boldsymbol{x}}$ and $\mathcal{L}^{\boldsymbol{y}}$ differ is at least $k$.
          Next, we will show that this contradicts $A+B+C \equiv 0 \pmod{tP}$. We first bound the values of $A$ and $C$. Assume $\sum_{\ell=i}^{i+t-1} x_\ell= \lambda$ where $0 \leq \lambda \leq t-1$, then $-\lambda(t-1-\lambda) \leq A \leq \lambda(t-\lambda)$, and the minimum and maximum values can be achieved when $(\boldsymbol{x}_{[i,i+t-1]}, \boldsymbol{y}_{[i,i+t-2]})= (1^{\lambda} 0^{t-\lambda}, 0^{t-1-\lambda} 1^{\lambda})$ and $(\boldsymbol{x}_{[i,i+t-1]}, \boldsymbol{y}_{[i,i+t-2]})= (0^{t-\lambda} 1^{\lambda}, 1^{\lambda} 0^{t-1-\lambda})$, respectively. Similarly, assume $\sum_{\ell=j}^{j+t-1} y_{\ell}= \lambda'$, we obtain $-\lambda'(t-1-\lambda') \leq C \leq \lambda'(t-\lambda')$.
          Then $|A+B+C|< \lambda(t-\lambda) + \lambda'(t-\lambda') + P \leq \frac{t^2}{2} + P < tP$.
          If $A+B+C \equiv 0 \pmod{tP}$ then it must be the case that $B=-A-C\leq \lambda(t-1-\lambda)+ \lambda'(t-1-\lambda')$, i.e., the number of ones in $\boldsymbol{x}_{[i+t,j]}$ is at most $\lambda(t-1-\lambda)+ \lambda'(t-1-\lambda')$. As a result the number of ones in $\boldsymbol{x}_{[i,j+t-1]}$ is at most $\lambda(t-\lambda)+ \lambda'(t-\lambda') \leq \lfloor \frac{t^2}{2} \rfloor= k$, which indicates that the distance between the leftmost and rightmost indices on which $\mathcal{L}^{\boldsymbol{x}}$ and $\mathcal{L}^{\boldsymbol{y}}$ differ is at most $k-1$.
          Thus, $A+B+C \equiv 0 \pmod{tP}$ does not hold.

      \item $\Delta \in [1,t-1]$. Assume $\sum_{\ell=i}^{i+t-1} x_{\ell}= \lambda$ and then $\sum_{\ell=i}^{i+t-2} y_{\ell}= \lambda + \Delta \leq t-1$. Now we bound the values of $A$ and $C$. It holds that $-\lambda (t-\lambda-1) -\sum_{\ell=i+t-1-\lambda-\Delta}^{i+t-\lambda-2} \ell \leq A \leq \lambda (t-\lambda)- \sum_{\ell=i+\lambda}^{i+\lambda+\Delta-1} \ell$, where the minimum and maximum values can be achieved when $(\boldsymbol{x}_{[i,i+t-1]}, \boldsymbol{y}_{[i,i+t-2]})= (1^{\lambda} 0^{t-\lambda}, 0^{t-1-\lambda-\Delta} 1^{\lambda+\Delta})$ and $(\boldsymbol{x}_{[i,i+t-1]}, \boldsymbol{y}_{[i,i+t-2]})= (0^{t-\lambda} 1^{\lambda}, 1^{\lambda+\Delta} 0^{t-1-\lambda-\Delta})$, respectively. Similarly, assume $\sum_{\ell=j}^{j+t-1} y_{\ell} = \lambda'$, we obtain $\sum_{\ell=j+1}^{j+t-1} x_{\ell}= \lambda'+\Delta \leq t-1$ and $-\lambda' (t-\lambda'-1)+ \sum_{\ell=j+\lambda'+1}^{j+\lambda'+\Delta} \ell \leq C \leq \lambda' (t-\lambda')+ \sum_{\ell=j+t-\lambda'-\Delta}^{j+t-\lambda'-1} \ell$. Now, we bound the value of $A+B+C$. On one hand, we have
          \begin{align*}
            &A+B+C\\
            &\leq \left( \lambda (t-\lambda)- \sum_{\ell=i+\lambda}^{i+\lambda+\Delta-1} \ell \right) + (P-2t+1) \\
            &~~~~~~~~+ \left( \lambda' (t-\lambda')+ \sum_{\ell=j+t-\lambda'-\Delta}^{j+t-\lambda'-1} \ell \right) \\
            &= \lambda (t-\lambda) + \lambda' (t-\lambda') + (P-2t+1) \\
            &~~~~~~~~+ \Delta (j-i+t-\lambda-\lambda'-\Delta) \\
            & \leq t \lambda + t \lambda' + (P-2t+1) \\
            &~~~~~~~~+ (t-1)(j-i+t-\lambda-\lambda')
            \\
            &= \lambda + \lambda' + P-2t+1+ (t-1)(j-i+t) \\
            & \leq P-1 + (t-1)P \\
            &= tP-1.
          \end{align*}
          On the other hand, we have
          \begin{align*}
            &A+B+C\\
            & \geq \left( -\lambda (t-\lambda-1) -\sum_{\ell=i+t-1-\lambda-\Delta}^{i+t-\lambda-2} \ell \right) \\
            &~~~~~~~~+ \left( -\lambda' (t-\lambda'-1)+ \sum_{\ell=j+\lambda'+1}^{j+\lambda'+\Delta} \ell \right) \\
            &= -\lambda (t-\lambda-1)- \lambda' (t-\lambda'-1) \\
            &~~~~~~~~+ \Delta (j+\lambda'+\Delta-i-t+\lambda+2) \\
            & \geq -\lambda (t-\lambda-1)- \lambda' (t-\lambda'-1) \\
            &~~~~~~~~+ (j+\lambda'-i-t+\lambda+3) \\
            & \geq -\lambda (t-\lambda-2)- \lambda' (t-\lambda'-2) + (j-i-t+3) \\
            & \geq - \lfloor \tfrac{1}{2} (t-2)^2 \rfloor + j-i-t+3 >0,
          \end{align*}
          where the last inequality holds since $j+t-1-i \geq k = \lfloor \frac{1}{2} t^2\rfloor$. Thus, $A+B+C \equiv 0 \pmod{tP}$ does not hold.

      \item $\Delta \in [-t,-2]$. Recall that we have $\Delta= \sum_{\ell=i}^{i+t-2}y_{\ell}- \sum_{\ell=i}^{i+t-1} x_{\ell}= \sum_{\ell=j+1}^{j+t-1}x_{\ell}- \sum_{\ell=j}^{j+t-1} y_{\ell}$, then $\sum_{\ell=i}^{i+t-2}(1-y_{\ell})- \sum_{\ell=i}^{i+t-1} (1-x_{\ell})= \sum_{\ell=j+1}^{j+t-1}(1-x_{\ell})- \sum_{\ell=j}^{j+t-1} (1-y_{\ell})= -1-\Delta \in [1,t-1]$.
          Consider the two sequences $\overline{\boldsymbol{x}}$ and $\overline{\boldsymbol{y}}$, then $\mathrm{VT}(\overline{\boldsymbol{x}})= \binom{n+1}{2}- \mathrm{VT}(\boldsymbol{x}) \equiv  \binom{n+1}{2}- \mathrm{VT}(\boldsymbol{y}) = \mathrm{VT}(\overline{\boldsymbol{y}}) \pmod{tP}$. 
          In a similar way as we do for the case when $\Delta \in [1,t-1]$, we obtain a contradiction.
      \item $\Delta = -1$. 
      Recall that $\Delta= \sum_{\ell=i}^{i+t-2}y_{\ell}- \sum_{\ell=i}^{i+t-1} x_{\ell}= \sum_{\ell=j+1}^{j+t-1}x_{\ell}- \sum_{\ell=j}^{j+t-1} y_{\ell}$, then we have $\sum_{\ell=i}^{i+t-2}(1-y_{\ell})- \sum_{\ell=i}^{i+t-1} (1-x_{\ell})= \sum_{\ell=j+1}^{j+t-1}(1-x_{\ell})- \sum_{\ell=j}^{j+t-1} (1-y_{\ell})= -1-\Delta= 0$.
      Again consider the two sequences $\overline{\boldsymbol{x}}$ and $\overline{\boldsymbol{y}}$, then $\mathrm{VT}(\overline{\boldsymbol{x}}) \equiv \mathrm{VT}(\overline{\boldsymbol{y}}) \pmod{tP}$ and $\mathrm{Sum}(A_k(\mathcal{L}^{\overline{\boldsymbol{x}}})_j) \equiv \mathrm{Sum}(A_k(\mathcal{L}^{\overline{\boldsymbol{y}}})_j) \pmod{t}$ for $j \in [1,k]$. In a similar way as we do for the case when $\Delta= 0$, we obtain a contradiction.

\end{itemize}

    In summary, the code $\mathcal{C}_{t,t-1}(n,P,2;\boldsymbol{a},\boldsymbol{b},\boldsymbol{c},\boldsymbol{c}')$ is a $P$-bounded binary $(t,t-1)$-burst correcting code. Then by the pigeonhole principle, there exists a choice of parameters, such that the code $\mathcal{C}_{t,t-1}(n,P,2;\boldsymbol{a},\boldsymbol{b},\boldsymbol{c},\boldsymbol{c}')$ has at most
    \begin{equation*}
      \log (tP) + \log (2t) + k (\log 2 + 2 \log t)= \log P + O(1)
    \end{equation*}
    bits of redundancy.
\end{IEEEproof}

\begin{remark}
     By setting $P=n$, we observe that the code $\mathcal{C}_{t,t-1}(n,P,2;\boldsymbol{a},\boldsymbol{b},\boldsymbol{c},\boldsymbol{c}')$ is indeed a binary $(t,t-1)$-burst correcting code and we abbreviate it as $\mathcal{C}_{t,t-1}(n,2;\boldsymbol{a},\boldsymbol{b},\boldsymbol{c},\boldsymbol{c}')$. Moreover, there exists a choice of parameters, such that the code $\mathcal{C}_{t,t-1}(n,2;\boldsymbol{a},\boldsymbol{b},\boldsymbol{c},\boldsymbol{c}')$ has at most $\log n +O(1)$
    bits of redundancy.
\end{remark}

\subsection{Non-Binary $(t,t-1)$-Burst Correcting Codes}

We now move on to the non-binary alphabet.
When $t=1$, $q$-ary $(t,t-1)$-burst correcting codes are essentially $q$-ary single-deletion correcting codes and are well understood. 
One well-known code construction in this context was introduced by Tenengolts \cite{Tenengolts-84-IT-q_D} using the signature function, which transforms $q$-ary sequences into binary sequences with desirable properties. 
When $t\geq 2$, we observe that the signature function still exhibits favorable properties.
For example, it aids in establishing a connection between 
$q$-ary $(t,t-1)$-burst correcting codes and binary $(t+1,t)$-burst correcting codes, as illustrated below.

\begin{definition}
For any sequence $\boldsymbol{x} \in \Sigma_q^n$, define its \emph{signature} $\alpha(\boldsymbol{x}) \in \Sigma_2^{n}$ to represent the relative order between consecutive symbols in $\boldsymbol{x}$, where $\alpha(x)_1=1$, $\alpha(x)_i=1$ if $x_{i}\geq x_{i-1}$ and $\alpha(x)_i=0$ otherwise, for $i\in [2,n]$.
\end{definition}

Figure \ref{fig:signature} shows how the signature changes when a $(t,t-1)$-burst error occurs in a sequence.
Then we have the following observation.

\begin{figure*}[htbp]
\begin{align*}
    \boldsymbol{x}= \cdots x_{i-1} x_{i} \underline{ x_{i+1} \cdots x_{i+t}} x_{i+t+1} x_{i+t+2} \cdots
    & \longrightarrow  \boldsymbol{y}= \cdots x_{i-1} x_{i} \underline{  y_{i+1} \cdots y_{i+t-1}} x_{i+t+1} x_{i+t+2} \cdots \\
    \alpha(\boldsymbol{x})= \cdots  \alpha(x)_{i} \underline{\alpha(x)_{i+1} \cdots \alpha(x)_{i+t} \alpha(x)_{i+t+1}} \alpha(x)_{i+t+2} \cdots &\longrightarrow \alpha(\boldsymbol{y})= \cdots \alpha(x)_{i} \underline{\alpha(y)_{i+1} \cdots \alpha(y)_{i+t}} \alpha(x)_{i+t+2} \cdots
\end{align*}
\caption{Illustrations of the changes in $\boldsymbol{x}$ and $\alpha(\boldsymbol{x})$ when $\boldsymbol{x}$ suffers a $(t,t-1)$-burst error.}
\label{fig:signature}
\end{figure*}

\begin{observation}\label{obs:bridge}
Any perturbation of $t$ consecutive coordinates in $\boldsymbol{x}$ will perturb at most $(t+1)$ consecutive coordinates in its signature. Thus, $\alpha(\boldsymbol{x})$ suffers a $(t+1,t)$-burst if $\boldsymbol{x}$ suffers a $(t,t-1)$-burst.
\end{observation}

As a result, to construct a $q$-ary $(t,t-1)$-burst correcting code $\mathcal{C}$,
we have to make sure that $\{\alpha(\boldsymbol{x}):\boldsymbol{x}\in\mathcal{C}\}$, when considered as a set, is a binary $(t+1,t)$-burst correcting code $\mathcal{C}'$.
However, $\{\alpha(\boldsymbol{x}):\boldsymbol{x}\in\mathcal{C}\}$ is very likely to be a multi-set since many $q$-ary sequences may have the same signature. To build a $q$-ary $(t,t-1)$-burst correcting code $\mathcal{C}$ based on a binary $(t+1,t)$-burst correcting code $\mathcal{C}'$, we still have to make sure that for any two distinct sequences $\boldsymbol{x}$ and $\boldsymbol{y}$ in $\mathcal{C}$ with the same signature in $\mathcal{C}'$, it holds that $\mathcal{B}_{t,t-1}(\boldsymbol{x}) \cap \mathcal{B}_{t,t-1}(\mathcal{\boldsymbol{y}})=\varnothing$.
\begin{example}
    Let $t=3$, $n= 10$, $q=4$, $\boldsymbol{x}= 22131 23121$, $\boldsymbol{y}= 23121 23121 $, and $\boldsymbol{z}= 23121 22131$. It can be verified that $\alpha(\boldsymbol{x})= \alpha(\boldsymbol{y})= \alpha(\boldsymbol{z})= 11010 11010$.
    Moreover, we have
   \begin{itemize}
       \item $\mathcal{B}_{3,2}(\boldsymbol{x}) \cap \mathcal{B}_{3,2}(\boldsymbol{y}) \neq \varnothing$ since $2 \underline{213} 1 23121 \rightarrow 2 \underline{11} 1 23121$ and  $2 \underline{312} 1 23121 \rightarrow 2 \underline{11} 1 23121$;
       \item $\mathcal{B}_{3,2}(\boldsymbol{y}) \cap \mathcal{B}_{3,2}(\boldsymbol{z}) \neq \varnothing$ since $ 23121 2 \underline{312} 1 \rightarrow 23121 2 \underline{11} 1$ and $23121 2 \underline{213} 1 \rightarrow 23121 2 \underline{11} 1$;
       \item $\mathcal{B}_{3,2}(\boldsymbol{x}) \cap \mathcal{B}_{3,2}(\boldsymbol{z}) = \varnothing$.
   \end{itemize}
   These three sequences share the same signature, and only $\boldsymbol{x}$ and $\boldsymbol{z}$ can belong to a $q$-ary $(3,2)$-burst correcting code simultaneously.
\end{example}

We construct non-binary $(t,t-1)$-burst correcting codes as follows.
For any codeword $\boldsymbol{x}$, we first require that its signature $\alpha(\boldsymbol{x})$ belongs to some given binary $(t+1,t)$-burst correcting code, such that we can recover $\alpha(\boldsymbol{x})$ from $\alpha(\boldsymbol{z})$, when $\boldsymbol{x}$ suffers a $(t,t-1)$-burst and results in $\boldsymbol{z}$. Then we use some additional constant bits of redundancy to guarantee that $\boldsymbol{x}$ can be uniquely decoded based on $\alpha(\boldsymbol{x})$ and some other clues of the sequence $\boldsymbol{x}$ itself, where the clues can be derived based on some additional constraints of the code.
For the simplest scenario where $t = 1$, Tenengolts \cite{Tenengolts-84-IT-q_D} demonstrated that a straightforward sum constraint modulo $q$ on the sequences themselves is sufficient for this purpose. However, for more challenging scenarios where $t \geq 2$, there are currently no efficient solutions. In the following, we will identify two distinct types of sum constraints that, when applied together, can effectively achieve this objective.

Before outlining the additional constraints, we first analyze to what extent could we learn about a sequence $\boldsymbol{x}$ when we have decoded $\alpha(\boldsymbol{x})$ from $\alpha(\boldsymbol{z})$.
An important observation is that we can always locate all the erroneous coordinates of $\boldsymbol{x}$ to be near a maximal monotone segment (unless otherwise stated, monotone means {\it monotone nondecreasing} or {\it monotone strictly decreasing}). To explicitly formalize this observation we need the following definition, which records the starting coordinate of each maximal monotone segment of $\boldsymbol{x}$.

\begin{definition}
For $\boldsymbol{x} \in \Sigma_q^n$, append $x_0=0$ at the very beginning. Let $\boldsymbol{x}^i$ be the $i$-th maximal monotone segment of $(0,\boldsymbol {x})$ and $m$ be the number of maximal monotone segments in $(0,\boldsymbol {x})$. Define $S(\boldsymbol{x})= S(x)_1 S(x)_2 \cdots S(x)_{m} S(x)_{m+1}$, where $S(x)_1= 0$, $S(x)_{m+1}=n$, and $S(x)_i$ denotes the coordinate of the first symbol of $\boldsymbol{x}^i$ in $(0, \boldsymbol{x})$ for $i\in[2,m]$.
\end{definition}

\begin{remark}\label{rmk:monotone}
Note that each maximal monotone segment is of length at least two, and there is exactly one overlapping coordinate between two neighbouring maximal monotone segments, i.e., the last symbol of each maximal monotone segment is the first symbol of the next maximal monotone segment. Therefore, $S(x)_{i+1}-S(x)_i= |\boldsymbol{x}^i|-1 \geq 1$.
\end{remark}

\begin{remark}\label{rmk:signature}
Note that the signature of a monotone non-decreasing segment of length $\ell$ is $1^{\ell-1}$ and the signature of a monotone strictly decreasing segment of length $\ell$ is $0^{\ell-1}$, thus, $\alpha(\boldsymbol{x})= 1^{S(x)_2-S(x)_1} 0^{S(x)_3-S(x)_2} \cdots$. In other words, the signature $\alpha(\boldsymbol{x})$ could be derived from the sequence $S(\boldsymbol{x})$, and vice versa.
\end{remark}

\begin{example}
  Let $\boldsymbol x=132434412132 \in \Sigma_5^{12}$, then $\boldsymbol{x}^1=013,\boldsymbol{x}^2=32,\boldsymbol{x}^3=24,\boldsymbol{x}^4=43, \boldsymbol{x}^5=344, \boldsymbol{x}^6=41, \boldsymbol{x}^7=12, \boldsymbol{x}^8=21, \boldsymbol{x}^9=13, \boldsymbol{x}^{10}=32$, $S(\boldsymbol{x})=(0,2,3,4,5,7,8,9,10,11,12)$, and $\alpha(\boldsymbol{x})= 110101101010$.
\end{example}

\begin{lemma}\label{lem:capability}
  Assume $\boldsymbol{x}$ suffers a $(t,t-1)$-burst and results in $\boldsymbol{z}$. Once the signature $\alpha(\boldsymbol{x})$ is known, we can locate all the erroneous coordinates of $\boldsymbol{x}$ near a maximal monotone segment. More precisely, we can always find some integer $\ell$, such that all the erroneous coordinates in $\boldsymbol{x}$ belong to the interval $[S(x)_\ell-t+1,S(x)_{\ell+1}+t-1]$.
\end{lemma}

\begin{IEEEproof}
By Remark \ref{rmk:signature}, with the help of $\alpha(\boldsymbol{x})$, we can determine the sequence $S(x)$. The type of errors occurring in $\alpha(\boldsymbol{x})$ can be determined by comparing $\alpha(\boldsymbol{x})$ and $\alpha(\boldsymbol{z})$:
\begin{itemize}
  \item either $\alpha(\boldsymbol{z})$ is obtained from $\alpha(\boldsymbol{x})$ by a $(\tilde{t},\tilde{t}-1)$-exact-burst for some $\tilde{t} \in [2,t+1]$;
  \item or $\alpha(\boldsymbol{z})$ is obtained from $\alpha(\boldsymbol{x})$ by a single deletion.
\end{itemize}

For the first case, let $i$ be the leftmost coordinate on which $\alpha(\boldsymbol{x})$ and $\alpha(\boldsymbol{z})$ differ, then the erroneous coordinates in $\alpha(\boldsymbol{x})$
are $\{i,i+1,\ldots,j \triangleq i+\tilde{t}-1\}$ where $\alpha(x)_j \neq \alpha(z)_{j-1}$ and $\alpha(x)_k = \alpha(z)_{k-1}$ for $k>j$, as shown below. 
\begin{align*}
  &\cdots \alpha(x)_{i-1} \underline{\alpha(x)_i \cdots \alpha(x)_j} \alpha(x)_{j+1} \cdots \\ \longrightarrow &\cdots \alpha(x)_{i-1} \underline{\alpha(z)_{i} \cdots \alpha(z)_{j-1}} \alpha(x)_{j+1} \cdots
\end{align*}
According to the definition of the signature, the erroneous coordinates in $\boldsymbol{x}$ contain at least one of $\{i-1,i\}$, and one of $\{j-1,j\}$. Assume $i \in [S(x)_\ell,S(x)_{\ell+1}]$ for some appropriate $\ell$, then all the erroneous coordinates in $\boldsymbol{x}$ belong to the interval $[j-t,i+t-1] \subseteq [i-t+1,i+t-1] \subseteq [S(x)_\ell-t+1,S(x)_{\ell+1}+t-1]$.

For the second case, assume the single deletion occurs in the $\ell$-th run of $\alpha(\boldsymbol{x})$ (a run in a sequence is a maximal substring consisting of the same symbol). We claim that all the erroneous coordinates in $\boldsymbol{x}$ belong to the interval $[S(x)_\ell-t+1,S(x)_{\ell+1}+t-1]$. Otherwise, the subword $[S(x)_\ell,S(x)_{\ell+1}]$ of $\boldsymbol{x}$ is not distorted and the $\ell$-th run of $\alpha(\boldsymbol{x})$ remains unchanged, which leads to a contradiction.

In summary, for both types of errors, we can always find some integer $\ell$, such that all the erroneous coordinates in $\boldsymbol{x}$ belong to the interval $[S(x)_\ell-t+1,S(x)_{\ell+1}+t-1]$, which completes the proof.
\end{IEEEproof}

\begin{example}\label{ex:signature}
Set $n= 12$, $q=5$, and $t=2$. Pick $\boldsymbol{x}= 132434412132$, which is unknown to the decoder. Suppose the decoder has derived $\alpha(\boldsymbol{x})= 110101101010$. Then $S(x)_5= 5$, $S(x)_6=7$, $S(x)_7= 8$, and $S(x)_8=9$.
  \begin{itemize}
    \item If $\boldsymbol{x}$ suffers a $(2,1)$-burst at the $7$-th coordinate and results in $\boldsymbol{z}= 132434 \underline{3} 2132$, the decoder can calculate $\alpha(\boldsymbol{z})= 110101 \underline{00} 010$. By comparing $\alpha(\boldsymbol{x})$ and $\alpha(\boldsymbol{z})$, it can be found that $\alpha(\boldsymbol{x})$ suffers a $(3,2)$-exact-burst at the $7$-th coordinate ($7$ is the leftmost coordinate on which $\alpha(\boldsymbol{x})$ and $\alpha(\boldsymbol{z})$ differ). Observe that $7 \in [S(x)_5,S(x)_6]$. Then the decoder concludes that the erroneous coordinates can be located in the interval $[4,8]$ of $\boldsymbol{x}$. Thus, the decoder can determine $(\boldsymbol{x}_{[1,3]}, \boldsymbol{x}_{[9,12]}) = (\boldsymbol{z}_{[1,3]}, \boldsymbol{z}_{[8,11]})= (132,2132)$.

    \item If $\boldsymbol{x}$ suffers a $(2,1)$-burst at the $6$-th coordinate and results in $\boldsymbol{z}= 13243 \underline{2} 12132$, the decoder can calculate $\alpha(\boldsymbol{z})= 11010 \underline{0} 01010$. By comparing $\alpha(\boldsymbol{x})$ and $\alpha(\boldsymbol{z})$, it can be found that $\alpha(\boldsymbol{x})$ suffers a $(2,1)$-exact-burst at the $6$-th coordinate ($6$ is the leftmost coordinate on which $\alpha(\boldsymbol{x})$ and $\alpha(\boldsymbol{z})$ differ). Observe that $6 \in [S(x)_5,S(x)_6]$. Then the decoder concludes that the erroneous coordinates can be located in the interval $[4,8]$ of $\boldsymbol{x}$. Thus, the decoder can determine $(\boldsymbol{x}_{[1,3]}, \boldsymbol{x}_{[9,12]}) = (\boldsymbol{z}_{[1,3]}, \boldsymbol{z}_{[8,11]})= (132,2132)$.
    \item If $\boldsymbol{x}$ suffers a $(2,1)$-burst at the $7$-th coordinate and results in $\boldsymbol{z}= 132434 \underline{2} 2132$, the decoder can calculate $\alpha(\boldsymbol{z})= 11010 \underline{1} 01010$. By comparing $\alpha(\boldsymbol{x})$ and $\alpha(\boldsymbol{z})$, it can be found that $\alpha(\boldsymbol{x})$ suffers a single deletion at the $5$-th run. Then the decoder concludes that the erroneous coordinates can be located in the interval $[S(x)_5-1,S(x)_6+1]= [4,8]$ of $\boldsymbol{x}$. Thus, the decoder can determine $(\boldsymbol{x}_{[1,3]}, \boldsymbol{x}_{[9,12]}) = (\boldsymbol{z}_{[1,3]}, \boldsymbol{z}_{[8,11]})= (132,2132)$. 
    \item If $\boldsymbol{x}$ suffers a $(2,1)$-burst at the $7$-th coordinate and results in $\boldsymbol{z}= 132434 \underline{4} 2132$, the decoder can calculate $\alpha(\boldsymbol{z})= 1101011 \underline{00} 10$. By comparing $\alpha(\boldsymbol{x})$ and $\alpha(\boldsymbol{z})$, it can be found that $\alpha(\boldsymbol{x})$ suffers a single deletion at the $7$-th run. Then the decoder concludes that the erroneous coordinates can be located in the interval $[S(x)_7-1,S(x)_8+1]= [7,10]$ of $\boldsymbol{x}$. Thus, the decoder can determine $(\boldsymbol{x}_{[1,6]}, \boldsymbol{x}_{[11,12]}) = (\boldsymbol{z}_{[1,6]}, \boldsymbol{z}_{[10,11]})= (132434,32)$. 
  \end{itemize}
\end{example}

Based on Lemma \ref{lem:capability}, after decoding $\alpha(\boldsymbol{x})$ from $\alpha(\boldsymbol{z})$, the decoder can already determine a large proportion of the correct sequence except for $\boldsymbol{x}_{[S(x)_\ell-t+1,S(x)_{\ell+1}+t-1]}$. Recall that $S(x)_{i+1}-S(x)_i \geq 1$ (Remark \ref{rmk:monotone}), and thus we have $[S(x)_\ell-t+1,S(x)_{\ell+1}+t-1] \subseteq [S(x)_{\ell-t+1},S(x)_{\ell+t}]$. That is, the unknown subsequence has a maximum of $2t$ consecutive monotone segments. Moreover, the subsequence of $\boldsymbol{x}$ with indices in the interval $[S(x)_{\ell+2it}-t+1,S(x)_{\ell+2it+1}+t-1] \subseteq [S(x)_{\ell+(2i-1)t+1},S(x)_{\ell+(2i+1)t}]$ is error-free, for $i \neq 0$. These facts inspire us to treat any $2t$ consecutive monotone segments as a whole and implement two constraints using constant bits of redundancy to gradually recover the desired subsequence. Observe that $\boldsymbol{x}_{[S(x)_{j}-t+1,S(x)_{j+1}+t-1]}$, for $j\in [1,|S(\boldsymbol{x})|]$, can be divided into three parts:
\begin{itemize}
  \item a substring of length $(t-1)$ on the left of the $j$-th maximal monotone segment, denoted as $\boldsymbol{x}^{j,L} \triangleq \boldsymbol{x}_{[S(x)_{j}-t+1,S(x)_{j}-1]}\in \Sigma_q^{t-1}$ (recall that we set $x_i=0$ if $i<0$);
  \item the $j$-th maximal monotone segment $\boldsymbol{x}^{j}$ (recall that $\boldsymbol{x}^{j}= \boldsymbol{x}_{[S(x)_{j},S(x)_{j+1}]}$);
  \item a substring of length $(t-1)$ on the right of the $j$-th maximal monotone segment, denoted as $\boldsymbol{x}^{j,R} \triangleq \boldsymbol{x}_{[S(x)_{j+1}+1,S(x)_{j+1}+t-1]}\in \Sigma_q^{t-1}$ (recall that we set $x_i=0$ if $i>n$).
\end{itemize}

\subsubsection{The First Constraint}
Use coordinate-wise sum constraints on the multi-sets of substrings $\{\boldsymbol{x}^{j,L}:j\in [1,|S(\boldsymbol{x})|]\}$ and $\{\boldsymbol{x}^{j,R}:j\in [1,|S(\boldsymbol{x})|]\}$, respectively, to guarantee that $\boldsymbol{x}$ can be recovered with the exception of at most one monotone segment.

\begin{lemma}\label{lem:monotone-segment}
Assume $\boldsymbol{x} \in \Sigma_q^n$ satisfies the following two classes of constraints, where the sum of sequences of length $t-1$ indicates a coordinate-wise sum:
     \begin{itemize}
       \item $\sum_{i \geq 0} \boldsymbol{x}^{j+2ti,L} \equiv \boldsymbol{\gamma}_j \pmod{q}$ for $j \in [1,2t]$ where $\boldsymbol{\gamma}_j \in \Sigma_q^{t-1}$;
       \item $\sum_{i \geq 0} \boldsymbol{x}^{j+2ti,R} \equiv \boldsymbol{\gamma}_j' \pmod{q}$ for $j \in [1,2t]$ where $\boldsymbol{\gamma}_j' \in \Sigma_q^{t-1}$.
     \end{itemize}
If a $(t,t-1)$-burst occurs in the interval $[S(x)_\ell-t+1,S(x)_{\ell+1}+t-1]$ and the sequence $S(\boldsymbol{x})$ is known, then we can recover $\boldsymbol{x}^{\ell,L}$ and $\boldsymbol{x}^{\ell,R}$.
\end{lemma}

\begin{IEEEproof}
Recall that $\boldsymbol{x}_{[S(x)_{\ell+2it}-t+1,S(x)_{\ell+2it+1}+t-1]}$ is error free for $i \neq 0$, we can obtain the correct $\boldsymbol{x}^{\ell+2ti,L}$ and $\boldsymbol{x}^{\ell+2ti,R}$ with the help of $S(\boldsymbol{x})$ for $i \neq 0$. Assume $\ell \equiv j \pmod{2t}$, for some $j \in [1,2t]$, then $\boldsymbol{x}^{\ell,L}$ and $\boldsymbol{x}^{\ell,R}$ can be determined by
    \begin{equation*}
        \begin{cases}
            \boldsymbol{x}^{\ell,L}= \boldsymbol{\gamma}_j- \sum_{i \neq 0} \boldsymbol{x}^{\ell+2ti,L} \pmod{q}; \\
            \boldsymbol{x}^{\ell,R}= \boldsymbol{\gamma}_j'- \sum_{i \neq 0} \boldsymbol{x}^{\ell+2ti,R} \pmod{q}.
        \end{cases}
    \end{equation*}
\end{IEEEproof}

\begin{example}\label{ex:LR}
  Continuing with Example \ref{ex:signature}. The sequence $\boldsymbol{x}= 132434412132$ satisfies several constraints, which include $\sum_{i \geq 0} \boldsymbol{x}^{1+4i,L}=x_{-1}+x_4+x_9= 1\pmod{5}$ and $\sum_{i \geq 0} \boldsymbol{x}^{1+4i,R}=x_3+x_8+x_{12}= 0\pmod{5}$.

  Suppose $S(\boldsymbol{x})=(0,2,3,4,5,7,8,9,10,11)$ is known to the decoder. Assume $S(\boldsymbol{x})$ suffers a $(2,1)$-burst and the resultant sequence is $\boldsymbol{z}= 132434 \underline{3} 2132$. By the first case in Example \ref{ex:signature}, the decoder can locate all the erroneous coordinates in the interval $[S(x)_5-1,S(x)_6+1]= [4,8]$ (which implies that $\ell=5$) and determines $(\boldsymbol{x}_{[1,3]}, \boldsymbol{x}_{[9,12]})= (132,2132)$. Then $x_4$ can be decoded by $x_4=1-x_{-1}-x_9=4\pmod{5}$ and $x_8$ can be decoded by $x_8=0-x_1-x_{12}=1\pmod{5}$. Thus at this moment the decoder has determined $(\boldsymbol{x}_{[1,4]}, \boldsymbol{x}_{[8,12]})= (1324,12132)$, and the unknown part $\boldsymbol{x}_{[5,7]}$ is a monotone segment.
\end{example}

\subsubsection{The Second Constraint}
Use counting constraints on each symbol so as to determine the symbols in the unknown monotone segment and thus completely recover it.

\begin{lemma}\label{lem:variation2}
  Let $\boldsymbol{x} \in \Sigma_q^n$ satisfy $|\{j\in[n]:x_j= i \}| \equiv \beta_i \pmod{2t}$, where $\beta_{i} \in \mathbb{Z}_{2t}$ for $i \in [1,q-1]$,  if $\boldsymbol{x}$ suffers a $(t,t-1)$-burst and results in $\boldsymbol{z}$,
  we can determine the number of occurrences of each symbol in $\boldsymbol{x}$ from $\boldsymbol{z}$.
\end{lemma}

The proof is similar to Lemma \ref{lem:variation1} and is thus omitted.

For the unknown segment $\boldsymbol{x}^\ell$, we have retrieved the set of  symbols in this segment and we also know whether $\boldsymbol{x}^\ell$ is nondecreasing or strictly decreasing with the help of $\alpha(\boldsymbol{x})$.
Thus, there is exactly one way to arrange these symbols, which enables us to recover $\boldsymbol{x}^\ell$.

\begin{example}\label{ex:M}
  Continuing with Example \ref{ex:LR}. In sequence $\boldsymbol x=132434412132$, the number of occurrences of $0,1,2,3,4$ of $\boldsymbol{x}$ is $0,3,3,3,3$, respectively. Since $(\boldsymbol{x}_{[1,4]}, \boldsymbol{x}_{[8,12]})= (1324,12132)$ has been determined, then the decoder knows that the symbols in $\boldsymbol{x}_{[5,7]}$ should be $\{3,4,4\}$. Moreover, $\boldsymbol{x}_{[5,7]}$ is the 5th monotone segment and thus it is monotone non-decreasing, and thus $\boldsymbol{x}_{[5,7]}=344$.
\end{example}

The next theorem summarizes the additional constraints listed above and presents a construction of non-binary $(t,t-1)$-burst correcting codes.

\begin{theorem}\label{thm:q-ary(t,t-1)-burst}
     Set $k= \lfloor \frac{1}{2}(t+1)^2 \rfloor$. Let $\boldsymbol{a} \in \mathbb{Z}_{n(t+1)} \times \mathbb{Z}_{2(t+1)}$, $\boldsymbol{b} \in \mathbb{Z}_2^{k}$, and $\boldsymbol{c}, \boldsymbol{c}' \in \mathbb{Z}_{t+1}^k$. Let $\boldsymbol{\beta} \in \Sigma_{2t}^{q-1}$, and $\boldsymbol{\gamma}, \boldsymbol{\gamma}' \in \Sigma_{q}^{2t(t-1)}$.
    We define the code $\mathcal{C}_{t,t-1}(n,q; \boldsymbol{\beta},\boldsymbol{\gamma},\boldsymbol{\gamma}') \subseteq \Sigma_q^n$ in which each sequence $\boldsymbol{x}$ satisfies the following constraints:
  \begin{itemize}
    \item $|\{j: x_j= i\}| \equiv \beta_{i} \pmod{2t}, ~ \forall 1 \leq i \leq q-1$;
    \item $\sum_{i \geq 0} \boldsymbol{x}^{j+2ti,L} \equiv \boldsymbol{\gamma}_{[(j-1)(t-1)+1,j(t-1)]} \pmod{q}$ for all $j \in [1,2t]$;
    \item $\sum_{i \geq 0} \boldsymbol{x}^{j+2ti,R} \equiv \boldsymbol{\gamma}_{[(j-1)(t-1)+1,j(t-1)]}' \pmod{q}$ for all $j \in [1,2t]$.
  \end{itemize}
  Then the code $\mathcal{C}_{t,t-1}(n,q;\boldsymbol{a},\boldsymbol{b},\boldsymbol{c},\boldsymbol{c}'; \boldsymbol{\beta},\boldsymbol{\gamma},\boldsymbol{\gamma}') \triangleq \{\boldsymbol{x} \in \mathcal{C}_{t,t-1}(n,q; \boldsymbol{\beta},\boldsymbol{\gamma},\boldsymbol{\gamma}'):  \alpha(\boldsymbol{x}) \in \mathcal{C}_{t+1,t}(n,2;\boldsymbol{a},\boldsymbol{b},\boldsymbol{c},\boldsymbol{c}')\}$ is a $q$-ary $(t,t-1)$-burst correcting code. By the pigeonhole principle, there exists a choice of parameters such that its redundancy is at most $\log n+O(1)$.
\end{theorem}

\begin{IEEEproof}
If $\boldsymbol{x} \in \mathcal{C}_{t,t-1}(n,q;\boldsymbol{a},\boldsymbol{b},\boldsymbol{c},\boldsymbol{c}'; \boldsymbol{\beta},\boldsymbol{\gamma},\boldsymbol{\gamma}')$ suffers a $(t,t-1)$-burst, we can recover its signature $\alpha(\boldsymbol{x})$ since $\alpha(\boldsymbol{x})$ belongs to a binary $(t+1,t)$-burst correcting code.
From $\alpha(\boldsymbol{x})$, $S(\boldsymbol{x})$ can be derived. By Lemma \ref{lem:capability}, we can locate all the erroneous coordinates in the interval $[S(x)_\ell-t+1, S(x)_{\ell+1}+t-1]$ for some $\ell$. It remains to recover $\boldsymbol{x}^{\ell,L}$, $\boldsymbol{x}^\ell$, and $\boldsymbol{x}^{\ell,R}$.
By Lemma \ref{lem:monotone-segment}, $\boldsymbol{x}^{\ell,L}$ and $\boldsymbol{x}^{\ell,R}$ can be recovered. Then by Lemma \ref{lem:variation2}, we can determine the number of occurrences of each symbol in $\boldsymbol{x}$ and uniquely
recover $\boldsymbol{x}^\ell$.
Thus, $\mathcal{C}_{t,t-1}(n,q;\boldsymbol{a},\boldsymbol{b},\boldsymbol{c},\boldsymbol{c}'; \boldsymbol{\beta},\boldsymbol{\gamma},\boldsymbol{\gamma}')$ is indeed a $q$-ary $(t,t-1)$-burst correcting code.
The redundancy can be calculated by the pigeonhole principle directly.
\end{IEEEproof}

In addition we have the $P$-bounded version for such codes.

\begin{corollary}\label{cor:q-ary(t,t-1)-burst}
     Set $k= \lfloor \frac{1}{2}(t+1)^2 \rfloor$. Let $\boldsymbol{a} \in \mathbb{Z}_{(P+1)(t+1)} \times \mathbb{Z}_{2(t+1)}$, $\boldsymbol{b} \in \mathbb{Z}_2^{k}$, and $\boldsymbol{c}, \boldsymbol{c}' \in \mathbb{Z}_{t+1}^k$. Let $\boldsymbol{\beta} \in \Sigma_{2t}^{q-1}$, and $\boldsymbol{\gamma}, \boldsymbol{\gamma}' \in \Sigma_{q}^{2t(t-1)} $. The code $\mathcal{C}_{t,t-1}(n,P,q;\boldsymbol{a},\boldsymbol{b},\boldsymbol{c},\boldsymbol{c}'; \boldsymbol{\beta},\boldsymbol{\gamma},\boldsymbol{\gamma}') \triangleq \{ \boldsymbol{x} \in \mathcal{C}_{t,t-1}(n,q; \boldsymbol{\beta},\boldsymbol{\gamma},\boldsymbol{\gamma}'): \alpha(\boldsymbol{x}) \in \mathcal{C}_{t+1,t}(n,P+1,2;\boldsymbol{a},\boldsymbol{b},\boldsymbol{c},\boldsymbol{c}')
    \big\}$ is a $P$-bounded $q$-ary $(t,t-1)$-burst correcting code. Then by the pigeonhole principle, there exists a choice of parameters, such that its redundancy is at most $\log P+O(1)$.
\end{corollary}

\begin{IEEEproof}
    If $\boldsymbol{x} \in \mathcal{C}_{t,t-1}(n,P,q;\boldsymbol{a},\boldsymbol{b},\boldsymbol{c},\boldsymbol{c}'; \boldsymbol{\beta},\boldsymbol{\gamma},\boldsymbol{\gamma}')$ suffers a $(t,t-1)$-burst with the erroneous coordinates located within a known  interval of length $P$, by Observation \ref{obs:bridge}, $\alpha(\boldsymbol{x})$ suffers a $(t+1,t)$-burst with the erroneous coordinates located in an interval of length $P+1$.
    We can then recover its signature $\alpha(\boldsymbol{x})$ since $\alpha(\boldsymbol{x})$ belongs to a binary $(P+1)$-bounded $(t+1,t)$-burst correcting code.
    The remainder of the proof follows the same logic as that of Theorem \ref{thm:q-ary(t,t-1)-burst}.
\end{IEEEproof}

\subsection{$(t,s)$-Burst Correcting Codes}

Recall that in Section \ref{Sec:Con}, for any $t>s$, we establish the connection between $(t,s)$-burst correcting codes and $(t',t'-1)$-burst correcting codes where $t'= \lceil t/(t-s) \rceil+1$. In Theorem \ref{thm:q-ary(t,t-1)-burst}, we have constructed $(t',t'-1)$-burst correcting codes. Then by Theorem \ref{thm:ET_ts}, we obtain the following theorem immediately. Here we assume $(t-s)|n$, otherwise we append zeros at the very end of each sequence such that its length is the smallest integer greater than $n$ and divisible by $(t-s)$.

\begin{theorem}\label{thm:(t,s)-burst}
  Suppose $t>s$. Set $t'= \lceil t/(t-s) \rceil+1$ and $k= \lfloor \frac{1}{2}(t+1)^2 \rfloor$. Let $\boldsymbol{a} \in \mathbb{Z}_{\frac{n}{t-s}(t'+1)} \times \mathbb{Z}_{2(t'+1)}$, $\boldsymbol{b} \in \mathbb{Z}_2^{k}$, and $\boldsymbol{c}, \boldsymbol{c}' \in \mathbb{Z}_{t'+1}^k$. Let $\boldsymbol{\beta} \in \Sigma_{2t'}^{q^{t-s}-1}$, and $\boldsymbol{\gamma}, \boldsymbol{\gamma}' \in \Sigma_{q^{t-s}}^{2t'(t'-1)}$. Assume $\mathcal{C}_{t',t'-1}(\frac{n}{t-s},q^{t-s};\boldsymbol{a},\boldsymbol{b},\boldsymbol{c},\boldsymbol{c}'; \boldsymbol{\beta},\boldsymbol{\gamma},\boldsymbol{\gamma}')$ is a $q^{t-s}$-ary  $(t',t'-1)$-burst correcting code constructed in Theorem \ref{thm:q-ary(t,t-1)-burst}, then the code
  \begin{equation*}
  \begin{aligned}
    &\mathcal{C}_{t,s}(n,q;\boldsymbol{a},\boldsymbol{b},\boldsymbol{c},\boldsymbol{c}'; \boldsymbol{\beta},\boldsymbol{\gamma},\boldsymbol{\gamma}') \triangleq \Big\{\boldsymbol{x} \in \Sigma_q^n: \\
    &~~~~\mathcal{D}_{t-s}(\boldsymbol{x}) \in \mathcal{C}_{t',t'-1}(\tfrac{n}{t-s},q^{t-s};\boldsymbol{a},\boldsymbol{b},\boldsymbol{c},\boldsymbol{c}'; \boldsymbol{\beta},\boldsymbol{\gamma},\boldsymbol{\gamma}') \Big\}
  \end{aligned}
  \end{equation*}
  is a $q$-ary $(t,s)$-burst correcting code. Moreover, there exists a choice of parameters such that the corresponding redundancy is at most $\log n + O(1)$.
\end{theorem}

Note that if $\boldsymbol{x}$ suffers a $(t,s)$-burst and we have found an interval of length $P$ which contains all the erroneous coordinates of $\boldsymbol{x}$, then $\mathcal{D}_{t-s}(\boldsymbol{x})$ suffers a $(t',t'-1)$-burst and we are able to find an interval of length $\lceil P/(t-s) \rceil+1$, which contains all the erroneous coordinates of $\mathcal{D}_{t-s}(\boldsymbol{x})$. Thus, we obtain the following corollary.

\begin{corollary}\label{cor:q-ary(t,s)-burst}
Set $P' \triangleq \lceil P/(t-s) \rceil+1$, $t'= \lceil t/(t-s) \rceil+1$, and $k= \lfloor \frac{1}{2}(t'+1)^2 \rfloor$. Let $\boldsymbol{a} \in \mathbb{Z}_{(P'+1) (t'+1)} \times \mathbb{Z}_{2(t'+1)}$, $\boldsymbol{b} \in \mathbb{Z}_2^{k}$, and $\boldsymbol{c}, \boldsymbol{c}' \in \mathbb{Z}_{t'+1}^k$. Let $\boldsymbol{\beta} \in \Sigma_{2t'}^{q^{t-s}-1}$, and $\boldsymbol{\gamma}, \boldsymbol{\gamma}' \in \Sigma_{q^{t-s}}^{2t'(t'-1)} $.
Assume $\mathcal{C}_{t',t'-1}(\tfrac{n}{t-s},P',q^{t-s};\boldsymbol{a},\boldsymbol{b},\boldsymbol{c},\boldsymbol{c}'; \boldsymbol{\beta},\boldsymbol{\gamma},\boldsymbol{\gamma}')$ is a $q^{t-s}$-ary $P'$-bounded $(t',t'-1)$-burst correcting code constructed in Corollary \ref{cor:q-ary(t,t-1)-burst}, then the code 
  \begin{equation*}
  \begin{aligned}
     &\mathcal{C}_{t,s}(n,P,q;\boldsymbol{a},\boldsymbol{b},\boldsymbol{c},\boldsymbol{c}'; \boldsymbol{\beta},\boldsymbol{\gamma},\boldsymbol{\gamma}')\triangleq \Big\{\boldsymbol{x} \in \Sigma_q^n: \\
     &~~~~\mathcal{D}_{t-s}(\boldsymbol{x}) \in \mathcal{C}_{t',t'-1}(\tfrac{n}{t-s},P',q^{t-s};\boldsymbol{a},\boldsymbol{b},\boldsymbol{c},\boldsymbol{c}'; \boldsymbol{\beta},\boldsymbol{\gamma},\boldsymbol{\gamma}') \Big\}
  \end{aligned}
  \end{equation*}
  is a $q$-ary $P$-bounded $(t,s)$-burst correcting code. Moreover, there exists a choice of parameters such that the corresponding redundancy is at most $\log P + O(1)$.
\end{corollary}

\section{Applications}\label{Sec:App}

In this section, we discuss the applications of our ($P$-bounded) $(t,s)$-burst correcting codes, and improve the relevant results under certain parameters.

\subsection{$q$-Ary Codes Correcting an Inversion Error}

An inversion error is defined as follows. For any sequence $\boldsymbol{x} \in \Sigma_q^n$, we say that $\boldsymbol{x}$ suffers an inversion of length exactly $t$ if some substring of $\boldsymbol{x}$ of length $t$, where the first and last entries are distinct, is reversed. For example, let $\boldsymbol{x}=\underline{0123}01230123$, then the sequence $\underline{3210} 0123 0123$ can be obtained when an inversion of length exactly $4$ occurs in $\boldsymbol{x}$. Similarly, we say that $\boldsymbol{x}$ suffers an inversion of length at most $t$ if some substring of $\boldsymbol{x}$ of length at most $t$ is reversed. A code $\mathcal{C}$ is called a $t$-inversion correcting code if it can correct an inversion of length exactly $t$, and called a $^{\leq}t$-inversion correcting code if it can correct an inversion of length at most $t$.

Motivated by applications in storing information in the live DNA, Nguyen \textit{et al}. \cite{Nguyen-22-ISIT} initiated the study of inversion correcting codes. They constructed $q$-ary $t$-inversion correcting codes and $^{\leq}t$-inversion correcting codes with $\log n + 4t \log q + 2$ and $\log n + 7(t + 1) \log q + 4$ bits of redundancy, respectively. We observe that an inversion of length exactly or at most $t$ can be seen a special case of a $(t,t)$-burst error. Therefore, by applying Theorem \ref{thm:(t,t)-burst}, the redundancy can be reduced to $\log n + 4(t-1) \log q + 2 -\log (t-1)$.

\subsection{$q$-Ary Codes Correcting an Absorption Error}

In \cite{Ye-24-IT-Absorption}, Ye \emph{et al.} proposed two types of absorption errors.
To distinguish between these two types of errors, we use the terms Type-A-Absorption error and Type-B-Absorption error and define them as follows.

Given a sequence $\boldsymbol{x} \in \Sigma_q^n$,
a Type-A-Absorption error at the $n$-th coordinate is simply a deletion of the $n$-th coordinate,
and a Type-A-Absorption error at the $i$-th coordinate, where $i\in [1,n-1]$, will delete $x_i$ and replace $x_{i+1}$ as $\min\{x_i+x_{i+1},q-1\}$.
For example, assume $\boldsymbol{x}= 01231113 \in \Sigma_4^8$.
If a Type-A-Absorption error occurs at the $6$-th, $7$-th, or the $8$-th coordinate, then the resultant sequence is $0123123$, $0123113$, and $0123111$, accordingly.
We say that $\mathcal{C} \subseteq \Sigma_q^n$ is a Type-A-Absorption correcting code if it can correct a Type-A-Absorption error.

Given a sequence $\boldsymbol{x} \in \Sigma_q^n$,
a Type-B-Absorption error at the $n$-th coordinate replaces $x_n$ as some value $x'_n$ such that $0\leq x'_n < x_n$.
A Type-B-Absorption error at the $i$-th coordinate, where $i\in [1,n-1]$, will replace $x_i$ as some value $x'_i$ such that $0\leq x'_i<x_i$,
and replace $x_{i+1}$ as $\min\{x_{i+1}+x_i-x'_i,q-1\}$.
For example, assume $\boldsymbol{x}= 01231312 \in \Sigma_4^8$.
If $\boldsymbol{x}$ suffers a Type-B-Absorption error at the $8$-th coordinate and $x_8$ is substituted as $0$ or $1$, the resultant sequence is $01231310$ or $01231311$, respectively.
If $\boldsymbol{x}$ suffers a Type-B-Absorption error at the $4$-th coordinate and $x_4$ is substituted as $0$, $1$, or $2$, the resultant sequence is $01203312$, $01213312$, and $01222312$, accordingly.
We say that $\mathcal{C} \subseteq \Sigma_q^n$ is a Type-B-Absorption correcting code if it can correct a Type-B-Absorption error.

Motivated by applications in neural and chemical communication systems, Ye \emph{et al.} \cite{Ye-24-IT-Absorption} initiated the study of absorption correcting codes, but only the Type-A-Absorption error was studied. They showed that the optimal redundancy of any Type-A-Absorption correcting code is lower bounded by $\log n + O(1)$ and gave a construction with $\log n + 12 \log \log n + O(1)$ bits of redundancy. We observe that a Type-A-Absorption error is a special case of $(2,1)$-burst error and a Type-B-Absorption error is a special case of $(2,2)$-burst error. Therefore, by applying Theorems \ref{thm:(t,t)-burst} and \ref{thm:(t,s)-burst}, we can construct absorption correcting codes for both Type-A and Type-B models with $\log n+O(1)$ bits of redundancy.

\subsection{$q$-Ary Codes Correcting a $^{\leq} t$-Burst-Deletion Error}

A $^{\leq}t$-burst-deletion correcting code can correct a burst of at most $t$ consecutive deletions. Intuitively, such a code can be constructed by using the intersection of $(i,0)$-burst correcting codes for all $i\in [1,t]$. However, this demands $t \log n$ bits of redundancy. In the following, we first introduce a lemma that will help us to find a small interval containing all the erroneous coordinates, and then use our $P$-bounded $(i,0)$-burst correcting codes to reduce the overall redundancy.

\begin{lemma}\cite[Lemma 2]{Lenz-20-ISIT-BD}\label{lem:loc code}
There exists a code $\mathcal{C}_1 \subseteq \Sigma_2^n$ with $\log n + O(1)$ bits of redundancy, such that the decoder can locate all the erroneous coordinates in an interval of length at most $O(\log n)$, if  $\boldsymbol{x} \in \mathcal{C}_1$ suffers a burst of at most $t$ consecutive deletions.
\end{lemma}

To consider the $q$-ary alphabet, we give the following indicator vector.

\begin{definition}
    For any $\boldsymbol{x} \in \Sigma_q^n$, we define its indicator vector $\mathbbm{1}(\boldsymbol{x}) \in \Sigma_2^n$ where $\mathbbm{1}(x)_i= 1$ if $x_i$ is odd and $\mathbbm{1}(x)_i=0$ otherwise.
\end{definition}

\begin{theorem}\label{thm:leq_t-burst}
    Set $P=O(\log n)$, $t'= 2$, and $k= \lfloor \frac{1}{2}(t'+1)^2 \rfloor$. Let $P_i= \lceil P/i \rceil +1$, $\boldsymbol{a}_i \in \mathbb{Z}_{(P_i+1) (t'+1)} \times \mathbb{Z}_{2(t'+1)}$, $\boldsymbol{b}_i \in \mathbb{Z}_2^{k}$, and $\boldsymbol{c}_i, \boldsymbol{c}_i' \in \mathbb{Z}_{t'+1}^k$, for all $i \in [1,t]$. Let $\boldsymbol{\beta}_i \in \Sigma_{2t'}^{q^i-1}$, and $\boldsymbol{\gamma}_i, \boldsymbol{\gamma}_i' \in \Sigma_{q^i}^{2t'(t'-1)}$, for all $i \in [1,t]$. Assume $\mathcal{C}_{i,0}(n,P,q;\boldsymbol{a}_i,\boldsymbol{b}_i,\boldsymbol{c}_i,\boldsymbol{c}_i'; \boldsymbol{\beta}_i,\boldsymbol{\gamma}_i,\boldsymbol{\gamma}'_i)$ is a $q$-ary $P$-bounded $(i,0)$-burst correcting code constructed in Corollary \ref{cor:q-ary(t,s)-burst}, then the code
    \begin{align*}
        &\Big\{ \boldsymbol{x} \in \Sigma_q^n: \mathbbm{1}(\boldsymbol{x}) \in \mathcal{C}_1, \\
        &~~~~\boldsymbol{x} \in \cap_{i=1}^t \mathcal{C}_{i,0}(n,P,q;\boldsymbol{a}_i,\boldsymbol{b}_i,\boldsymbol{c}_i,\boldsymbol{c}_i'; \boldsymbol{\beta}_i,\boldsymbol{\gamma}_i,\boldsymbol{\gamma}'_i) \Big\}
    \end{align*}
    is a $q$-ary $^{\leq}t$-burst-deletion correcting code. Moreover, when $q$ is even, there exists a choice of parameters, such that the corresponding redundancy is at most $\log n+ t\log P +O(1)= \log n+ t\log \log n +O(1)$.
\end{theorem}

\begin{IEEEproof}
    Assume $\boldsymbol{x}$ suffers a burst of at most $t$ consecutive deletions and the resultant sequence is $\boldsymbol{z}$, then the number of deletions can be determined by comparing the length of $\boldsymbol{x}$ and $\boldsymbol{z}$. Assume $\boldsymbol{x}$ suffers an $i$-burst deletion. On one hand, we can locate all the erroneous coordinates in an interval of length $P=O(\log n)$ since $\mathbbm{1}(\boldsymbol{x}) \in \mathcal{C}_1$. On the other hand, $\boldsymbol{x}$ belongs to some $P$-bounded $i$-burst-deletion correcting code and thus can be recovered. 

    To calculate the code size, we first note that by Lemma \ref{lem:loc code} there is a binary code $\mathcal{C}_1$ of size $\frac{2^n}{Cn}$ for some constant $C$. For each binary sequence, there are $(q/2)^n$ $q$-ary sequences sharing it as a common indicator vector, according to the definition of indicator vector. Therefore, by the pigeonhole principle, the code size is at least
    \begin{align*}
    \frac{(q/2)^n\cdot \frac{2^n}{Cn}}{\prod\limits_{i=1}^t\left( (P_i+1)2(t'+1)^2 2^{k}(t'+1)^{2k}(2t')^{q^i-1}(q^i)^{4t'(t'-1)} \right)},
    \end{align*}
    which implies that the corresponding redundancy is at most $\log n+t\log\log n+O(1)$.
\end{IEEEproof}

\begin{remark}
For even $q$, our code constructed in Theorem $\ref{thm:leq_t-burst}$ is better than the results of Schoney \emph{et al.} \cite{Schoeny-17-IT-BD} and Lenz \emph{et al.} \cite{Lenz-20-ISIT-BD}, which have redundancy $\log n+ \binom{t+1}{2}\log \log n+O(1)$. Moreover, when $t< 8\lceil \log q \rceil$, our code constructed in Theorem $\ref{thm:leq_t-burst}$ is also better than Wang \emph{et al.} \cite{Wang-22-arXiv-BD}, which has redundancy $\log n+ 8 \lceil \log q \rceil \log \log n+O(1)$.
\end{remark}

\subsection{$q$-Ary Codes Correcting a $t$-Localized-Deletion Error}

We say that $\boldsymbol{x}$ suffers a burst of at most $t$ non-consecutive deletions, or a $t$-localized-deletion, if it suffers $i$ deletions for some $i \leq t$, where the deleted coordinates are localized in an interval of length at most $t$. A $t$-localized-deletion correcting code can correct a $t$-localized-deletion. Note that the deletion of some $i \leq t$ coordinates in an interval of length $t$ can be seen as a special case of a $(t,t-i)$-burst error. Thus, $t$-localized-deletion correcting codes can be constructed by using the intersection of $(t,t-i)$-burst correcting codes for all $i \in [1,t]$. However, this incurs $t \log n$ bits of redundancy. Similar to the construction of $^{\leq}t$-burst-deletion correcting codes, we first introduce a lemma that will help us to find a small interval containing all the erroneous coordinates, and then use our $P$-bounded $(t,t-i)$-burst correcting codes to reduce the overall redundancy.

\begin{lemma}\cite[Section III.C]{Bitar-21-ISIT-LD}
    There exists a code $\mathcal{C}_2 \subseteq \Sigma_2^n$ with $\log n +O(1)$ bits of redundancy, such that the decoder can locate all the erroneous coordinates in an interval of length at most $O(\log^2 n)$, if $\boldsymbol{x} \in \mathcal{C}_2$ suffers a burst of at most $t$ non-consecutive deletions.
\end{lemma}

\begin{theorem}\label{thm:localized_deletion}
    Set $P= O(\log^2 n)$. Let $P_i= \lceil P/i \rceil+ 1$, $t_i'= \lceil t/i \rceil+1$, $k_i= \lfloor \frac{1}{2}(t_i'+1)^2 \rfloor$, $\boldsymbol{a}_i \in \mathbb{Z}_{(P_i+1)(t_i'+1)} \times \mathbb{Z}_{2(t_i'+1)}$, $\boldsymbol{b}_i \in \mathbb{Z}_2^{k_i}$, and $\boldsymbol{c}_i, \boldsymbol{c}_i' \in \mathbb{Z}_{t_i'+1}^{k_i}$, for all $i \in [1,t]$. Let $\boldsymbol{\beta}_i \in \Sigma_{2t_i'}^{q^i-1}$, and $\boldsymbol{\gamma}_i, \boldsymbol{\gamma}_i' \in \Sigma_{q^{i }}^{2t_i'(t_i'-1)}$, for all $i \in [1,t]$.
    Assume $\mathcal{C}_{t,t-i}(n,P,q;\boldsymbol{a}_i,\boldsymbol{b}_i,\boldsymbol{c}_i,\boldsymbol{c}_i'; \boldsymbol{\beta}_i,\boldsymbol{\gamma}_i,\boldsymbol{\gamma}'_i)$ is a $q$-ary $P$-bounded $(t,t-i)$-burst correcting code constructed in Corollary \ref{cor:q-ary(t,s)-burst}, then the code
    \begin{align*}
        &\Big\{ \boldsymbol{x} \in \Sigma_q^n: \mathbbm{1}(\boldsymbol{x}) \in \mathcal{C}_2, \\
        &~~~~\boldsymbol{x} \in \cap_{i=1}^t \mathcal{C}_{t,t-i}(n,P,q;\boldsymbol{a}_i,\boldsymbol{b}_i,\boldsymbol{c}_i,\boldsymbol{c}_i'; \boldsymbol{\beta}_i,\boldsymbol{\gamma}_i,\boldsymbol{\gamma}'_i) \Big\}
    \end{align*}
    is a $t$-localized-deletion correcting code. Moreover, when $q$ is even, there exists a choice of parameters, such that the corresponding redundancy is at most $\log n + 2t\log\log n+O(1)$.
\end{theorem}

\begin{IEEEproof}
    Assume $\boldsymbol{x}$ suffers a burst of at most $t$ non-consecutive deletions and the resultant sequence is $\boldsymbol{z}$, then the number of deletions can be determined by comparing the length of $\boldsymbol{x}$ and $\boldsymbol{z}$, says $i$. That is, $\boldsymbol{x}$ suffers a $(t,t-i)$-burst. On one hand, we can locate all the erroneous coordinates in an interval of length $P=O(\log^2 n)$ since $\mathbbm{1}(\boldsymbol{x}) \in \mathcal{C}_2$. On the other hand, $\boldsymbol{x}$ belongs to some $P$-bounded $(t,t-i)$-burst correcting code and thus can be recovered. The redundancy can be calculated in a manner similar to that of Theorem \ref{thm:leq_t-burst}.
\end{IEEEproof}

\begin{remark}
For even $q$, the code constructed in Theorem $\ref{thm:localized_deletion}$ is better than the code of Bitar \emph{et al.} \cite{Bitar-21-ISIT-LD}, which has redundancy $\log n+ 16t\log \log n +O(1)$.
\end{remark}

\subsection{Permutation Codes Correcting a $t$-Burst-Stable-Deletion Error}

Now, we turn our attention to a permutation code $\mathcal{C} \subseteq S_n$.
In permutations there are both stable and unstable deletions \cite{Gabrys-16-IT-PD}.
Here, the term of a stable deletion in $S_n$ is the same as a deletion in $\Sigma_q^n$. We say that $\mathcal{C} \subseteq S_n$ is a $t$-BSD correcting permutation code if it can correct a burst of $t$ consecutive stable deletions.

In permutation codes, the size of the alphabet equals the codeword length, which makes designing low redundancy error correcting codes challenging. Previous works \cite{Chee-20-IT-PBSD,Sun-22-IT-BD,Wang-22-arXiv-BD} proposed several techniques to reduce the alphabet size. A connection between $t$-BSD correcting permutation codes and $(t+1)!$-ary $(2t,t)$-burst correcting codes is implicit in \cite{Wang-22-arXiv-BD}. Thus, we can use our $(2t,t)$-burst correcting codes to improve the corresponding redundancy. Firstly, we introduce the ideas from \cite{Wang-22-arXiv-BD}.

\begin{definition}\label{def:permutation_projection}
Let $\boldsymbol{u} = (u_1,u_2,\ldots,u_n)$ be a vector with integer entries. We define $\mathrm{Prj}(\boldsymbol{u}) \in S_n$ be the \emph{permutation projection} of $\boldsymbol{u}$, where $\mathrm{Prj}(u)_i = | \left\{j : u_j < u_i,1 \leq j \leq n \right\}| + | \left\{j : u_j = u_i,1 \leq j \leq i \right\}|$ for all $i\in [1,n]$.
\end{definition}

\begin{definition}\label{def:permutation_rank}
    Let $\mu: \mathcal{S}_n \rightarrow[1, n !]$ be a bijection such that $\mu(\boldsymbol{\sigma})$ is the \emph{lexicographic rank} of $\boldsymbol{\sigma}$ in $\mathcal{S}_n$. For any sequence $\boldsymbol{u}$ of length $n$, we define its \emph{permutation rank} as $r(\boldsymbol{u})=\mu(\mathrm{Prj}(\boldsymbol{u}))$.
\end{definition}

\begin{definition}\label{def:overlapping_ranking_sequence}
 For any permutation $\boldsymbol{\sigma} \in S_n$, we define $\boldsymbol{p}_{t+1}(\boldsymbol{\sigma})= (p_1, p_2,\ldots, p_{n-t}) \in \Sigma_{(t+1)!}^{n-t}$ as its \emph{$t$-overlapping ranking sequence} where $p_i$ is the permutation rank of $(\sigma_i,\sigma_{i+1},\ldots,\sigma_{i+t})$ for all $i\in [1,n-t]$.
\end{definition}

\begin{lemma}\cite[Claim 1]{Wang-22-arXiv-BD}\label{lem:(2t,t)-burst}
For any permutation $\boldsymbol{\sigma} \in S_n$, if $\boldsymbol{\sigma}$ suffers a burst of $t$ consecutive stable deletions, the corresponding $t$-overlapping ranking sequence $\boldsymbol{p}_{t+1}(\boldsymbol{\sigma})$ suffers a $t$-burst-substitution followed by a $t$-burst-deletion, which is thus a special case of a $(2t,t)$-burst error.
\end{lemma}

\begin{lemma}\cite[Lemma 12]{Wang-22-arXiv-BD}\label{lem:overlapping_ranking_sequence}
    For two permutations $\boldsymbol{\sigma}$ and $\boldsymbol{\pi}$ in $S_n$, if they share a subsequence of length $n-t$ which can be obtained from each of them after a burst of $t$ consecutive stable deletions, then $\boldsymbol{p}_{t+1}(\boldsymbol{\sigma}) \neq \boldsymbol{p}_{t+1}(\boldsymbol{\pi})$.
\end{lemma}

Lemma \ref{lem:overlapping_ranking_sequence} suggests that if we want to correct a burst of $t$ consecutive stable deletions occurring in a permutation $\boldsymbol{\sigma} \in S_n$, it suffices to recover its $t$-overlapping ranking sequence $\boldsymbol{p}_{t+1}(\boldsymbol{\sigma})$. Then by Lemma \ref{lem:(2t,t)-burst}, we know that $\boldsymbol{p}_{t+1}(\boldsymbol{\sigma}) \in \Sigma_{(t+1)!}^{n-t}$ suffers a $(2t,t)$-burst if $\boldsymbol{\sigma}$ suffers a burst of $t$ consecutive stable deletions. Thus, we can construct $t$-BSD correcting permutation codes by using $(t+1)!$-ary $(2t,t)$-burst correcting codes.

\begin{theorem}\label{thm:t-BSD_permutation_code}
    Set $q=(t+1)!$, $t'= \lceil 2t/t \rceil+1= 3$, and $k= \lfloor \frac{1}{2}(t'+1)^2 \rfloor$.  Let $\boldsymbol{a} \in \mathbb{Z}_{\frac{n-t}{t}(t'+1)} \times \mathbb{Z}_{2(t'+1)}$, $\boldsymbol{b} \in \mathbb{Z}_2^{k}$, and $\boldsymbol{c}, \boldsymbol{c}' \in \mathbb{Z}_{t'+1}^k$. Let $\boldsymbol{\beta} \in \Sigma_{2t'}^{q^{t}-1}$, and $\boldsymbol{\gamma}, \boldsymbol{\gamma}' \in \Sigma_{q^{t}}^{2t'(t'-1)}$. Assume $\mathcal{C}_{2t,t}(n-t,q;\boldsymbol{a},\boldsymbol{b},\boldsymbol{c},\boldsymbol{c}'; \boldsymbol{\beta},\boldsymbol{\gamma},\boldsymbol{\gamma}')$ is a $q$-ary  $(2t,t)$-burst correcting code constructed in Theorem \ref{thm:q-ary(t,t-1)-burst}, then the code
  \begin{equation*}
  \begin{aligned}
    \Big\{\boldsymbol{\sigma} \in S_n: \boldsymbol{p}_t(\boldsymbol{\sigma}) \in \mathcal{C}_{2t,t}(n-t,q;\boldsymbol{a},\boldsymbol{b},\boldsymbol{c},\boldsymbol{c}'; \boldsymbol{\beta},\boldsymbol{\gamma},\boldsymbol{\gamma}') \Big\},
  \end{aligned}
  \end{equation*}
  is a $t$-BSD correcting permutation code. Moreover, there exists a choice of parameters such that the corresponding redundancy is at most $\log n + O(1)$.
\end{theorem}

\begin{IEEEproof}
    Assume $\boldsymbol{\sigma}$ suffers a burst of $t$ consecutive stable deletions and the resultant sequence is $\boldsymbol{z}$, then by Lemma \ref{lem:(2t,t)-burst}, we have that $\boldsymbol{p}_{t+1}(\boldsymbol{z}) \in \mathcal{B}_{2t,t}(\boldsymbol{p}_{t+1}(\boldsymbol{\sigma}))$. Since $\boldsymbol{p}_{t+1}(\boldsymbol{\sigma})$ belongs to some $(2t,t)$-burst correcting code, we can recover $\boldsymbol{p}_{t+1}(\boldsymbol{\sigma})$ from $\boldsymbol{p}_{t+1}(\boldsymbol{z})$. Now, by Lemma \ref{lem:overlapping_ranking_sequence}, we can further recover $\boldsymbol{\sigma}$ with the help of $\boldsymbol{z}$ and $\boldsymbol{p}_{t+1}(\boldsymbol{\sigma})$. The redundancy can be calculated by the pigeonhole principle directly.
\end{IEEEproof}

\begin{remark}
The code constructed in Theorem $\ref{thm:t-BSD_permutation_code}$ is better than the code of Sun \emph{et al.} \cite{Sun-22-IT-BD}, which has redundancy $\log n+ 2\log \log n +O(1)$.
\end{remark}

\subsection{Permutation Codes Correcting a $^{\leq}t$-Burst-Deletion Error}

We say that $\mathcal{C} \subseteq S_n$ is a $^{\leq}t$-BSD correcting permutation code if it can correct a burst of at most $t$ consecutive stable deletions. Obviously, $^{\leq}t$-BSD permutation correcting codes can be constructed by using the intersection of $i$-BSD correcting permutation codes for all $i\in [1,t]$. However, this incurs $t \log n$ bits of redundancy. 
Similar to the construction of $^{\leq}t$-burst-deletion correcting codes, we first introduce a lemma that will help us to find a small interval containing all the erroneous coordinates, and then use our $P$-bounded $(2i,i)$-burst correcting codes to reduce the overall redundancy.

\begin{lemma}\cite[Lemmas 10 and 11]{Wang-22-arXiv-BD}
    There exists a code $\mathcal{C}_3 \subseteq S_n$ with $\log n +O(1)$ bits of redundancy, such that the decoder can locate all the erroneous coordinates in an interval of length at most $O(\log n)$, if $\boldsymbol{\sigma} \in \mathcal{C}_3$ suffers a burst of at most $t$ consecutive stable deletions.
\end{lemma}

Since $\boldsymbol{p}_{t+1}(\boldsymbol{\sigma})$ is based on the permutation rank of $(t+1)$ consecutive symbols, once we locate the errors of $\boldsymbol{\sigma}$ in an interval of length $P$, we are able to locate the errors in $\boldsymbol{p}_{t+1}(\boldsymbol{\sigma})$ in an interval of length $P+t$. Now, we are ready to present our $^{\leq}t$-BSD correcting permutation codes.

\begin{theorem}\label{thm:leqt-BSD_permutation_code}
Set $P=O(\log n)$, $t'= 3$, and $k= \lfloor \frac{1}{2}(t'+1)^2 \rfloor$.
Let $q_i=(i+1)!$, $P_i= P+i$, $P_i'= \lceil P_i/i \rceil+ 1$, $\boldsymbol{a}_i \in \mathbb{Z}_{(P_i'+1) (t'+1)} \times \mathbb{Z}_{2(t'+1)}$, $\boldsymbol{b}_i \in \mathbb{Z}_2^{k}$, and $\boldsymbol{c}_i, \boldsymbol{c}_i' \in \mathbb{Z}_{t'+1}^k$, for all $i\in [1,t]$. Let $\boldsymbol{\beta}_i \in \Sigma_{2t'}^{q^i-1}$, and $\boldsymbol{\gamma}_i, \boldsymbol{\gamma}_i' \in \Sigma_{q^i}^{2t'(t'-1)}$, for all $i\in [1,t]$.
Assume $\mathcal{C}_{2i,i}(n-i,P_i,q_i;\boldsymbol{a}_i,\boldsymbol{b}_i,\boldsymbol{c}_i,\boldsymbol{c}_i'; \boldsymbol{\beta}_i,\boldsymbol{\gamma}_i,\boldsymbol{\gamma}_i')$ is a $q_i$-ary $P_i$-bounded $(2i,i)$-burst correcting code constructed in Corollary \ref{cor:q-ary(t,t-1)-burst}, then the code
  \begin{equation*}
  \begin{aligned}
    &\Big\{\boldsymbol{\sigma} \in \mathcal{C}_3: \\
    &\boldsymbol{p}_{i+1}(\boldsymbol{\sigma}) \in \cap_{i=1}^t \mathcal{C}_{2i,i}(n-i,P_i,q_i;\boldsymbol{a}_i,\boldsymbol{b}_i,\boldsymbol{c}_i,\boldsymbol{c}_i'; \boldsymbol{\beta}_i,\boldsymbol{\gamma}_i,\boldsymbol{\gamma}_i') \Big\},
  \end{aligned}
  \end{equation*}
  is a $^{\leq}t$-BSD correcting permutation code. Moreover, there exists a choice of parameters such that the corresponding redundancy is at most $\log n + t \log \log n + O(1)$.
\end{theorem}

\begin{IEEEproof}
    Assume $\boldsymbol{\sigma}$ suffers a burst of at most $t$ consecutive stable deletions and the resultant sequence is $\boldsymbol{z}$, then the number of deletions can be determined by comparing the length of $\boldsymbol{\sigma}$ and $\boldsymbol{z}$, says $i$. Then by Lemma \ref{lem:(2t,t)-burst}, we have that $\boldsymbol{p}_{i+1}(\boldsymbol{z}) \in \mathcal{B}_{2i,i}(\boldsymbol{p}_{i+1}(\boldsymbol{\sigma}))$.
    On one hand, we can locate all the erroneous coordinates of $\boldsymbol{\sigma}$ in an interval of length $P$ since $\boldsymbol{\sigma} \in \mathcal{C}_3$, and further locate all the erroneous coordinates of $\boldsymbol{p}_{i+1}(\boldsymbol{\sigma})$ in an interval of length $P_i=P+i$. On the other hand, $\boldsymbol{p}_{i+1}(\boldsymbol{\sigma})$ belongs to some $P_i$-bounded $(2i,i)$-burst correcting code. Thus, we can recover $\boldsymbol{p}_{i+1}(\boldsymbol{\sigma})$, and further recover $\boldsymbol{\sigma}$. The redundancy can be calculated by the pigeonhole principle directly.
\end{IEEEproof}

\begin{remark}
The code constructed in Theorem $\ref{thm:leqt-BSD_permutation_code}$ is better than the code of Sun \emph{et al.} \cite{Sun-22-IT-BD}, which has redundancy $\log n+ (3t-2)\log \log n +O(1)$.
\end{remark}

\section{Conclusion}\label{Sec:Concl}

Motivated by applications in DNA storage, we consider codes against a burst of insertions, deletions, and substitutions. Our main contribution is to present explicit constructions of $(t,s)$-burst correcting codes with $\log n +O(1)$ redundant bits for any alphabet size $q$ and any constants $t$ and $s$. The gap between the redundancy of these codes and the theoretic bound is only $O(1)$. In addition, we use these codes or their $P$-bounded versions to cope with other types of errors and improve the redundancy under certain parameters. It would be interesting to find further applications of ($P$-bounded) $(t,s)$-burst correcting codes.

\end{document}